\makeatletter
\def\@cons#1#2{\begingroup\let\@elt\relax\xdef#1{\ifx#1\relax\else#1\fi\@elt #2}\endgroup}
\makeatother
\documentclass[twocolumn]{autart} 

\usepackage[dvipsnames]{xcolor}
\usepackage{cite}

\let\labelindent\relax
\usepackage{graphics} 
\usepackage{epsfig} 
\usepackage{amsmath} 
\newcommand\numberthis{\addtocounter{equation}{1}\tag{\theequation}}
\usepackage{amssymb} 
\usepackage{mathtools}
\usepackage{xspace}
\usepackage{algpseudocode}
\usepackage{centernot}
\usepackage{algorithm}
\usepackage{epstopdf}
\usepackage{tcolorbox}
\usepackage{varwidth}
\usepackage{bbm}
\usepackage{bm}
\usepackage{verbatim}

\let\theoremstyle\relax
\usepackage{amsthm}
\newtheoremstyle{spaced}  
  {6pt} 
  { } 
  {\itshape}     
  { }     
  {\bfseries} 
  {.}    
  { }    
  {}     
\theoremstyle{spaced}

\usepackage{siunitx}
\usepackage{cancel}

\usepackage{enumitem}

\usepackage{tikz} 
\usepgflibrary{arrows}

\usepackage[hypertexnames=false]{hyperref}
\usepackage[labelformat=simple]{subcaption}

\usepackage[normalem]{ulem}
\usepackage{enumitem}

\makeatletter
\let\cl@part\undefined
\makeatletter

\usepackage[capitalise]{cleveref} 

\crefname{assumption}{Assumption}{Assumptions}

\newcounter{relctr} 
\everydisplay\expandafter{\the\everydisplay\setcounter{relctr}{0}} 
\newcommand\labelrel[2]{%
  \begingroup
    \refstepcounter{relctr}%
    \stackrel{\textnormal{(\text{\roman{relctr})}}}{\mathstrut{#1}}%
    \originallabel{#2}%
  \endgroup
}
\AtBeginDocument{\let\originallabel\label} 

\newcommand{\manish}[1]{\textcolor{red}{Manish: #1}}

\newcommand{\JK}[1]{\textcolor{blue}{Johannes: #1}}

\allowdisplaybreaks
 
\newcommand{\Kinf}{\mathcal{K}_{\infty}}
\newcommand{\KL}{\mathcal{KL}}

\newcommand{\Lipstagecost}{L_{\ell}}
\newcommand{\Liptermcost}{L_{\mathrm{f}}}
\newcommand{\Intrange}[2]{\mathbb{Z}^{+}_{[#1,#2]}}

\newcommand{\Ljc}{L_c}

\newcommand{\coninput}{u}
\newcommand{\state}{x}
\newcommand{\gpinp}{z}
\newcommand{\dynGP}{g}
\newcommand{\dynGPtrue}{g^{\star}}
\newcommand{\dynNom}{f}
\newcommand{\dyntrue}{f^{\star}}
\newcommand{\statedim}{{n_\state}}
\newcommand{\inputdim}{{n_\coninput}}
\newcommand{\gpdim}{{n_\dynGP}}
\newcommand{\epsmpc}[1][]{\varepsilon_{{#1}}}
\newcommand{\epsdyn}{\varepsilon}
\newcommand{\epstightening}[1][]{\xi_{{#1}}}
\newcommand{\Lipdyn}{L}
\newcommand{\noise}{w}
\newcommand{\noisebound}{\bar{w}}
\newcommand{\datasetdim}{D}

\newcommand{\horizon}{H}
 
\newcommand{\numsamples}{N}

\newcommand{\n}{n}
\newcommand{\integer}{\mathbb{Z}}
\newcommand{\identity}{I}
\newcommand{\inftynorm}[1][]{{\|#1\|}_\infty}
\newcommand{\euclidnorm}[1][]{{\|#1\|}}
\newcommand{\stagecost}[2]{\ell(#1,#2)}
\newcommand{\stagecostNA}{\ell}
\newcommand{\terminalcost}[1]{V_{\mathrm{f}}(#1)}
\newcommand{\terminalcostNA}[1][]{V_{\mathrm{f}} #1}
\newcommand{\terminalinput}{\coninput_{\mathrm{f}}}
\newcommand{\terminalset}{\mathcal{X}_{\mathrm{f}}}
\newcommand{\GP}{\mathcal{GP}}

\newcommand{\kernelfunc}{k}

\newcommand{\gpsubspace}{B_d}
\newcommand{\Bg}{B_\dynGP}

\newcommand{\RKHS}{{\mathcal{H}_k}}

\newcommand{\RKHSnorm}[1][]{{\|#1\|}_k}
\newcommand{\RKHSdatanorm}[1][]{{\|#1\|}_{k_\mathrm{D}}}
\newcommand{\probability}[1]{\mathrm{Pr}\left( {#1} \right)}
\newcommand{\expo}[1]{e^{#1}}

\newcommand{\gppostvar}{\sigma}

\newcommand{\inputSpace}{\mathcal{U}}
\newcommand{\stateSpace}{\mathcal{X}}

\newcommand{\ki}{k}
\newcommand{\cost}{\ell}

\newcommand{\xpos}{x_p}
\newcommand{\ypos}{y_p}

\newcommand{\ball}[1]{\mathcal{B}_{#1}}

\newcommand{\mypar}[1]{\noindent\textbf{#1}.}

\renewcommand{\Re}{\mathbb{R}}
\newcommand{\Nat}{\mathbb{N}}
\newcommand{\Intp}{\mathbb{Z}^{+}}

\newcommand{\X}{\mathcal X}

\newcommand{\E}{\mathbb E}

\newcommand{\N}{\mathcal N}

\newcommand{\R}{\mathbb{R}}

\newtheorem{lemma}{Lemma}
\newtheorem{assumption}{Assumption}

\newtheorem{definition}{Definition}
\newtheorem{theorem*}{Theorem}
\newtheorem{corollary}{Corollary}

\newtheorem{theorem}{Theorem}
\newtheorem{remark}{Remark}

\newcommand{\muconst}[1][]{\mu_{#1}}

\newcommand{\betadata}[1][]{\beta_{#1}}

\input{v_line_patch}

\begin{document}

\begin{frontmatter}

\title{Finite-Sample-Based Reachability for Safe Control with \\ 
Gaussian Process Dynamics}


\thanks[footnoteinfo]{$\dagger$ The work is jointly supervised by Andreas Krause and Melanie N. Zeilinger. 
Manish Prajapat is supported by ETH AI center, Johannes K\"ohler by the Swiss National Science Foundation under NCCR Automation, grant agreement 51NF40 180545, and Amon Lahr by the European Union’s Horizon 2020 research and innovation programme, Marie Skłodowska-Curie grant agreement No. 953348, ELO-X.}

\author[ai_center,idsc,eth_cs]{Manish Prajapat}\ead{manishp@ai.ethz.ch},
\author[idsc]{Johannes K\"ohler},
\author[idsc]{Amon Lahr},    
\author[ai_center,eth_cs]{Andreas Krause$^\dagger$}  
\author[ai_center,idsc]{Melanie N. Zeilinger$^\dagger$},               

\address[ai_center]{ETH AI Center, ETH Zurich, Switzerland}  
\address[idsc]{Institute for Dynamic Systems and Control, ETH Zurich, Switzerland}             
\address[eth_cs]{Department of Computer Science, ETH Zurich, Switzerland}



\begin{keyword} 
Gaussian processes, 
Non-linear predictive control, 
Optimal controller synthesis for systems with uncertainties, 
Randomized methods, 
Robust control of nonlinear systems,
Learning theory, 
Sampling-based algorithm
\end{keyword}                             



\begin{abstract}
\looseness -1 

Gaussian Process~(GP) regression is shown to be effective for learning unknown dynamics, enabling efficient and safety-aware control strategies across diverse applications.
However, existing GP-based model predictive control (GP-MPC) methods either rely on approximations, thus lacking guarantees, or are overly conservative, which limits their practical utility. 
To close this gap, we present a sampling-based framework that efficiently propagates the model's epistemic uncertainty while avoiding conservatism.
We establish a novel sample complexity result that enables the construction of a reachable set using a finite number of dynamics functions sampled from the GP posterior. 
Building on this, we design a sampling-based GP-MPC scheme that is recursively feasible and guarantees closed-loop safety and stability with high probability. 
Finally, we showcase the effectiveness of our method on two numerical examples, highlighting accurate reachable set over-approximation and safe closed-loop performance.
\end{abstract} 

\end{frontmatter}



\section{Introduction}


\looseness -1 
Designing reliable control strategies typically requires a known system model, but accurately modeling system dynamics is often challenging in practice.
In such cases, Gaussian Process (GP) models~\cite{rasmussen_gaussian_2006} have proven effective for learning system dynamics from data. Their non-parametric nature allows them to flexibly model a wide variety of unknown functions, even under challenging and uncertain conditions~\cite{prajapat2022near,frigola2015bayesian,puigjaner2025performance,Tuong2010gp_manipulator,deringer2021gaussian}.
Moreover, GP regression offers a principled approach to characterize epistemic uncertainty arising from limited data, providing well-understood (high-probability) error bounds~\cite{abbasi2013online,beta_srinivas}. As a result, they are successfully applied in Bayesian optimization~\cite{frazier2018tutorial}, experiment design~\cite{prajapatsubmodular}, safety-critical control~\cite{lederer2019uniform}, and model predictive control (MPC)~\cite{prajapat2025safe}.



Particularly in MPC, GPs can be used to model uncertain dynamics, and the epistemic uncertainty can be propagated over the prediction horizon to construct reachable sets. This naturally leads to a safety-aware control~\cite{ostafew_robust_2016,cao_gaussian_2017,hewing_cautious_2020}. 
However, sequentially propagating epistemic uncertainty over the prediction horizon is challenging and often requires approximations~\cite{scampicchio2025gaussian}.
Common techniques include linearization-based approximation of dynamics~\cite{hewing_cautious_2020,vaskov_friction-adaptive_2022} or moment matching~\cite{quinonero-candela_propagation_2003}. 
These approximations, while tractable, make it difficult to ensure safety for the resulting GP-MPC formulation (cf. overview in~\cite{scampicchio2025gaussian}).
To guarantee safe operation, robust MPC approaches~\cite{marruedo2002ISSrobustMPC,sasfi2023robust,Houska2019robustMPCbook} can be used 
based on the high-probability GP error bounds~\cite {abbasi2013online,beta_srinivas}. 
A corresponding robust GP-MPC approach has been proposed in \cite{koller_learning-based_2018} by sequentially over-approximating ellipsoids using the worst-case GP estimate at each time independently. Although this approach can ensure safety, it suffers from potentially prohibitive conservatism and lacks recursive feasibility guarantees.
Notably, this conservatism is inherent in the sequential propagation used in most robust MPC approaches~\cite{marruedo2002ISSrobustMPC,sasfi2023robust,Houska2019robustMPCbook}, which cannot account for the time-invariance of the epistemic uncertainty. 

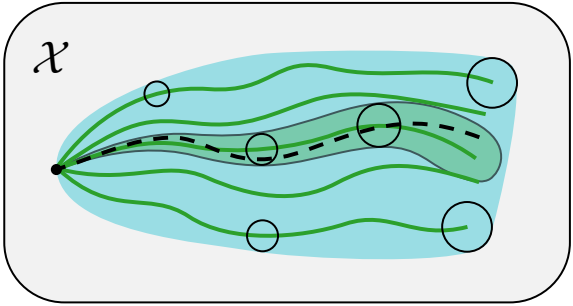
\begin{figure}
\centering
    \scalebox{1.0}{\tikzset{every picture/.style={line width=0.75pt}} 

\begin{tikzpicture}[x=0.75pt,y=0.75pt,yscale=-1,xscale=1]

\draw  [fill={rgb, 255:red, 128; green, 128; blue, 128 }  ,fill opacity=0.1 ] (122.73,102.07) .. controls (122.73,85.37) and (136.27,71.83) .. (152.97,71.83) -- (377.77,71.83) .. controls (394.46,71.83) and (408,85.37) .. (408,102.07) -- (408,192.77) .. controls (408,209.46) and (394.46,223) .. (377.77,223) -- (152.97,223) .. controls (136.27,223) and (122.73,209.46) .. (122.73,192.77) -- cycle ;
\draw  [draw opacity=0][fill={rgb, 255:red, 23; green, 190; blue, 207 }  ,fill opacity=0.4 ] (378.83,103.5) .. controls (387,112.67) and (371.83,177.5) .. (367.17,191.83) .. controls (362.5,206.17) and (268.83,200.83) .. (247.33,197) .. controls (225.83,193.17) and (150.2,188.29) .. (149,154.95) .. controls (147.8,121.6) and (227.4,99.2) .. (257.4,97.2) .. controls (287.4,95.2) and (370.67,94.33) .. (378.83,103.5) -- cycle ;
\draw  [color={rgb, 255:red, 0; green, 0; blue, 0 }  ,draw opacity=0.6 ][fill={rgb, 255:red, 44; green, 160; blue, 44 }  ,fill opacity=0.3 ][line width=0.75]  (359.67,132.17) .. controls (380.18,142.97) and (373,157.5) .. (367.5,160.83) .. controls (362,164.17) and (349,159.67) .. (337.83,152.17) .. controls (326.67,144.67) and (317.6,142) .. (293.2,147.2) .. controls (268.8,152.4) and (254.58,158.3) .. (218.69,148.95) .. controls (182.8,139.6) and (144.4,158.69) .. (149.2,155.95) .. controls (154,153.2) and (185.8,135.8) .. (216.75,138.09) .. controls (247.69,140.38) and (260.33,136.5) .. (289.17,127.33) .. controls (318,118.17) and (339.15,121.36) .. (359.67,132.17) -- cycle ;
\draw [color={rgb, 255:red, 44; green, 160; blue, 44 }  ,draw opacity=1 ][line width=1.5]    (149.2,154.95) .. controls (166.87,130.33) and (192.87,113) .. (225.87,115.33) .. controls (258.87,117.67) and (258.83,97.79) .. (287.53,104.67) .. controls (316.24,111.54) and (341.2,101.67) .. (369.2,112) ;
\draw [color={rgb, 255:red, 44; green, 160; blue, 44 }  ,draw opacity=1 ][line width=1.5]    (149.2,155.95) .. controls (186.2,130) and (194.53,130) .. (220.2,132.67) .. controls (245.87,135.33) and (257.53,127.67) .. (278.2,121) .. controls (298.87,114.33) and (326.53,119.67) .. (365.53,128) ;
\draw [color={rgb, 255:red, 44; green, 160; blue, 44 }  ,draw opacity=1 ][line width=1.5]    (149,155.95) .. controls (178.28,144.8) and (195,139.63) .. (217.62,144.19) .. controls (240.24,148.76) and (270.62,145.06) .. (289.69,137.92) .. controls (308.76,130.78) and (335.08,132.05) .. (360.57,150.31) ;
\draw [color={rgb, 255:red, 44; green, 160; blue, 44 }  ,draw opacity=1 ][line width=1.5]    (149.7,155.45) .. controls (177.53,181.33) and (195.2,166) .. (215.87,179.67) .. controls (236.53,193.33) and (272.03,190.67) .. (294.87,183.33) .. controls (317.7,176) and (328.99,191.87) .. (356.33,184.87) ;
\draw [color={rgb, 255:red, 44; green, 160; blue, 44 }  ,draw opacity=1 ][line width=1.5]    (149.2,155.45) .. controls (185.53,164.67) and (202.87,150.33) .. (224.87,160) .. controls (246.87,169.67) and (269.53,173.67) .. (289.87,163) .. controls (310.2,152.33) and (325.2,150.33) .. (362.2,162.33) ;
\draw   (193.43,117.75) .. controls (193.43,114.3) and (196.23,111.5) .. (199.68,111.5) .. controls (203.14,111.5) and (205.93,114.3) .. (205.93,117.75) .. controls (205.93,121.2) and (203.14,124) .. (199.68,124) .. controls (196.23,124) and (193.43,121.2) .. (193.43,117.75) -- cycle ;
\draw   (356.25,112.21) .. controls (356.25,105.28) and (361.87,99.67) .. (368.79,99.67) .. controls (375.72,99.67) and (381.33,105.28) .. (381.33,112.21) .. controls (381.33,119.13) and (375.72,124.75) .. (368.79,124.75) .. controls (361.87,124.75) and (356.25,119.13) .. (356.25,112.21) -- cycle ;
\draw   (343.58,184.87) .. controls (343.58,177.95) and (349.2,172.33) .. (356.13,172.33) .. controls (363.05,172.33) and (368.67,177.95) .. (368.67,184.87) .. controls (368.67,191.8) and (363.05,197.42) .. (356.13,197.42) .. controls (349.2,197.42) and (343.58,191.8) .. (343.58,184.87) -- cycle ;
\draw   (245.17,189.38) .. controls (245.17,185.07) and (248.67,181.57) .. (252.98,181.57) .. controls (257.3,181.57) and (260.8,185.07) .. (260.8,189.38) .. controls (260.8,193.7) and (257.3,197.2) .. (252.98,197.2) .. controls (248.67,197.2) and (245.17,193.7) .. (245.17,189.38) -- cycle ;
\draw [color={rgb, 255:red, 0; green, 0; blue, 0 }  ,draw opacity=1 ][line width=1.5]  [dash pattern={on 5.63pt off 4.5pt}]  (362.33,139.67) .. controls (330.67,127.83) and (313.67,132) .. (275.5,146) .. controls (237.33,160) and (226.8,138.83) .. (206,139.83) .. controls (185.2,140.83) and (175,147.5) .. (149,155.95) ;
\draw   (300.83,133.67) .. controls (300.83,127.68) and (305.68,122.83) .. (311.67,122.83) .. controls (317.65,122.83) and (322.5,127.68) .. (322.5,133.67) .. controls (322.5,139.65) and (317.65,144.5) .. (311.67,144.5) .. controls (305.68,144.5) and (300.83,139.65) .. (300.83,133.67) -- cycle ;
\draw   (244.77,145.78) .. controls (244.77,141.47) and (248.27,137.97) .. (252.58,137.97) .. controls (256.9,137.97) and (260.4,141.47) .. (260.4,145.78) .. controls (260.4,150.1) and (256.9,153.6) .. (252.58,153.6) .. controls (248.27,153.6) and (244.77,150.1) .. (244.77,145.78) -- cycle ;
\draw  [fill={rgb, 255:red, 0; green, 0; blue, 0 }  ,fill opacity=1 ] (146.71,155.95) .. controls (146.71,154.68) and (147.73,153.65) .. (149,153.65) .. controls (150.27,153.65) and (151.29,154.68) .. (151.29,155.95) .. controls (151.29,157.21) and (150.27,158.24) .. (149,158.24) .. controls (147.73,158.24) and (146.71,157.21) .. (146.71,155.95) -- cycle ;

\draw (135,90) node [anchor=north west][inner sep=0.75pt]  [font=\Large]  {$\mathcal{X}$};

\end{tikzpicture}}
    \caption{
    Illustration of the proposed sampling-based reachable set (blue region) based on a finite number of dynamics samples (green lines). 
    Using sampling, we treat the epistemic uncertainty to an arbitrarily small tolerance $\epsilon$ (i.e., at least one sample is $\epsilon$-close to the unknown dynamics). 
    We sequentially propagate the residual $\epsilon$-epistemic and aleatoric uncertainty, represented by black balls. As a result, the true (unknown) system response (black dashed line) is contained within a tube (green region) around the $\epsilon$-close sample.
    }
    \label{fig:reachability_explaination}
\end{figure}

\looseness -1 \emph{Sampling} is another widely used approach for reachability analysis, particularly in controls~\cite{lucia2015robust,calafiore2006scenario} and robotics~\cite{lew2022simple,williams2016aggressive}, which avoids the conservatism of sequential propagation techniques. 
To approximate the reachable set, sampling-based methods typically involve generating multiple realizations of uncertain model parameters and simulating the system dynamics for each sampled value~\cite{lucia2015robust,bonzanini_fast_2021}.
This leads to an inner approximation of the reachable set, which asymptotically converges to the true reachable set as the number of samples tends to infinity~\cite{lew2021sampling}.     
However, guarantees with finite samples are required to ensure safety while being computationally tractable. 
Given a known distribution over uncertain model parameters, statistical bounds for sampling-based methods can be computed using conformal prediction~\cite{lindemann2024formal}; however, these bounds are typically obtained post-execution, limiting their applicability in safety-critical settings.
Another sampling-based approach is scenario optimization~\cite{calafiore2006scenario}, which offers probabilistic guarantees assuming that the samples and the ground truth follow the same distribution. It has been used to compute reachable sets~\cite{devonport2020estimating}, applied with GP models~\cite{umlauft_scenario-based_2018-1}, and incorporated into MPC design~\cite{Campi2019Scenario}.
However, applying scenario arguments to non-convex programs remains challenging, typically resulting only in a-posteriori guarantees~\cite{garatti2021risk,lefringhausen2024learning}. 
For general-purpose reachability with finite-dimensional feature space, \cite{lew2022simple} obtains an $\epsilon$-close over-approximation and provides finite-sample guarantees using random set theory, which scales exponentially with the feature dimension. However, extending such results to GP-based models -- where the uncertainty lies in an infinite-dimensional function space -- is non-trivial.

In this work, we construct a finite-sample-based reachable set for GP dynamics that contains the unknown system trajectory with a user-specified high probability.
Building on this, we design a sampling-based GP-MPC framework that draws 
dynamic functions from the GP posterior and optimizes a control sequence to satisfy constraints across all sampled dynamics jointly.
By directly sampling dynamics functions, the method avoids worst-case independent propagation and thus efficiently propagates epistemic uncertainty without being conservative.
Our prior work \cite{prajapat2024towards} provides a numerical method for solving sampling-based GP-MPC problems; 
however, it lacks finite-sample guarantees and is not recursively feasible. 
We address these limitations by proposing a recursively feasible sampling-based GP-MPC framework that provides safety and stability guarantees with finite samples, through the following key contributions: 

\mypar{Finite Samples from GP posterior}
Our method treats epistemic uncertainty up to an arbitrary tolerance \mbox{$\epsilon>0$} by directly sampling functions from the GP posterior, akin to Thomson sampling~\cite{thompson1933likelihood,vakili2021scalable}.
To determine how many such functions are needed, we leverage coverage arguments~\cite{li2001gaussianSBP,stolz1996some,van2011information} in the associated Reproducing Kernel Hilbert Space (RKHS). 
Specifically, we bound the number of posterior GP samples required to ensure that, with high probability, at least one function is uniformly $\epsilon$-close to the unknown dynamics -- yielding a sample complexity bound.
This result is of independent interest for sampling-based methods in infinite-dimensional spaces. (\cref{sec:finite_sample_complexity})

\mypar{Sampling-based reachable set}
Assuming bounded disturbances and Lipschitz continuity of the unknown dynamics, we sequentially propagate the remaining $\epsilon$-epistemic uncertainty and aleatoric noise around the finite set of samples to construct a reachable set (\cref{fig:reachability_explaination}). We prove that, with high probability, the resulting reachable set contains the unknown future system trajectory at all times. (\cref{sec:reachable_set})

\mypar{Closed-loop guarantees} 
Building upon the proposed reachable set, we design a sampling-based GP-MPC strategy.
Due to non-sequential uncertainty propagation by the sampling-based approaches, 
ensuring recursive feasibility is non-trivial. 
To address this, we remove falsified samples at each timestep during closed-loop execution. 
Finally, we establish safety and stability guarantees for the resulting closed-loop system. 
This methodology is also of independent interest for sampling-based approaches in the finite-dimensional case, where recursive feasibility guarantees are typically lacking. (\cref{sec:MPC})

\mypar{Simulations} 
We demonstrate the effectiveness of our algorithm using a car model, comparing it against the robust GP-MPC method proposed in \cite{koller_learning-based_2018}, and further showcase the safe closed-loop performance of our method with a pendulum example. (\cref{sec:simulation})

\textit{Notation:} The set of non-negative integers is given by $\Intp$. For any $i,j \in \integer^{+}, i<j$, we use short hand notation $\Intrange{i}{j}$ for a set of integers $\{ i,i+1, \hdots, j\}$. 
We denote by $\mathbb{B}$ a Banach space of functions $g:\mathbb{R}^n\rightarrow\mathbb{R}$ equipped with the supremum norm $\inftynorm[\dynGP] = \sup_{x\in \mathbb{R}^n}|\dynGP(x)|$. 
We denote $f(x) = \mathcal{O}(g(x))$ as $x \to 0$ if $\exists x_0 > 0, c>0 : \forall~0 < x < x_0, f(x) \leq c g(x)$. 
We use $\|x\|$ to denote the Euclidean norm of a vector $x \in \R^n$, and $I$ for the identity matrix of appropriate dimension.
The Minkowski sum and Pontryagin set difference for two sets $X, Y \subseteq \R^n$ are defined by $X \oplus Y \coloneqq \{ x+y \in \R^\n | x \in X, y \in Y \}$ and $X \ominus Y \coloneqq \{ z \in \R^\n | z+y \in X~ \forall y \in Y \}$, respectively.
$\probability{X \geq x}$ denotes the probability of $X \geq x$.
The expectation of a function $f(w)$ over a
random variable $w$ is denoted by $\E[f(w)]$ and the expectation conditioned on an event $A$ is given by $\E[f(w)|A]$.
A sequence $\{\noise(\ki)\}_{\ki \geq 0}$ is conditionally $\lambda$-sub-Gaussian for a fixed $\lambda \geq0$ if $\forall \ki \in \Intp, \forall s \in \R, \E[e^{s \noise(\ki)}|\{\noise(i)\}_{i=1}^{\ki-1},\{\state(i)\}_{i=1}^{\ki}] \leq \mathrm{exp}(\frac{s^2 \lambda^2}{2})$ \cite{abbasi2013online,ao2025stochastic}. 
A continuous function $\alpha: \Re_{\geq0} \to \Re_{\geq0}$ belong to class $\mathcal{K}$ if it is strictly increasing, satisfy $\alpha (0)=0$ and belongs to class $\Kinf$ if additionally $\alpha$ is unbounded. A continuous function $\beta: \Re_{\geq0} \times \Re_{\geq0} \to \Re_{\geq0}$ is said to be class $\KL$ if for a fixed $s$, $\beta(r,s)$ belongs to class $\mathcal{K}$ and for a fixed $r$, $\beta(r,s)$ is decreasing with respect to $s$, s.t., $\beta(r,s) \to 0$ as $s\to \infty$.
\section{Problem statement}
We consider the task of controlling a discrete-time, nonlinear dynamical system 
\begin{align}
\!\!\! \state(\ki+1)\! = \!\dynNom(\state(\ki), \!\coninput(\ki)) + \gpsubspace \!\left( \dynGPtrue(\state(\ki), \!\coninput(\ki)) \!+\! \noise(k) \right) \! \label{eq:system_dyn}
\end{align}
with state $\state(\ki) \in \R^\statedim$ and input $\coninput(\ki) \in \R^\inputdim$. Here, \mbox{$\dynNom: \mathbb{R}^{\statedim} \times \mathbb{R}^{\inputdim} \rightarrow \mathbb{R}^{\statedim}$} denotes the known part of the dynamics, e.g., derived from first principles, and \mbox{$\dynGPtrue: \mathbb{R}^{\statedim} \times \mathbb{R}^{\inputdim} \rightarrow \mathbb{R}^{\gpdim}$} is the unknown part modelled in a subspace defined by \mbox{$\gpsubspace \in \mathbb{R}^{\statedim \times \gpdim}$} with full column rank. 
The unknown part is estimated using a prior dataset \mbox{$\mathcal{D} \coloneqq \{ Z, Y \}$}, constructed from $\datasetdim$ data points collected at 
\mbox{$Z = [ z_1, \ldots, z_\datasetdim ]$}, where each \mbox{$z_i \doteq (\state_i, \coninput_i) \in \mathbb{R}^{d}$} with $d = n_x +n_u$. 
The corresponding measurements are given by \mbox{$Y = [ y_1, \ldots, y_\datasetdim ]$}, where each \mbox{$y_i = \dynGPtrue(z_i) + \noise_i$} is one-step prediction error, readily obtained from \eqref{eq:system_dyn}.
 For simplicity of exposition, the results in this paper are presented for the case of a scalar unknown dynamics component,
i.e., \mbox{$\gpdim = 1$}; see~\cref{rem:vector_valued} 
for an extension to the multivariate case. 
The disturbance $\noise_i \in \mathbb{R}^{n_g}$ is assumed to be conditionally $\lambda$-sub-Gaussian, 
\mbox{$\lambda > 0$}. 
This includes the special case of zero-mean noise bounded in $[-\lambda,\lambda]$ and i.i.d. Gaussian noise with variance bounded by $\lambda^2$~\cite{abbasi2013online,prajapat2025safe}.


\mypar{Gaussian process} We use Gaussian process regression to quantify 
the unknown dynamics~$\dynGPtrue$ via a data-driven model. 
Given the dataset $\mathcal{D}$, the posterior mean and covariance of unknown dynamics at test inputs $z$, $z^{\prime}$ are given by:
\begin{align}
    \mu(z) &= \bm{\alpha}^\top \bm{k}_\datasetdim(z), \label{eq:mean_update}\\
    \kernelfunc_D(z,z') &= \kernelfunc(z,z) - \bm{\kernelfunc}_\datasetdim^\top(z)  
    (K_D + \lambda^2 \identity)^{-1}
    \bm{\kernelfunc}_\datasetdim(z'), \label{eq:kernel_posterior_update}\\ 
    \gppostvar^2(z) &= \kernelfunc_D(z,z),
\end{align} 
where $\bm{\alpha} = [\alpha_1, \hdots, \alpha_D]^\top \coloneqq (K_D + \lambda^2 \identity)^{-1} Y$, $\bm{\kernelfunc}_\datasetdim(z) = [k(z_1,z), . . . , k(z_{\datasetdim}, z)]^{\top}$, $\kernelfunc: \mathbb{R}^{n_z} \times \mathbb{R}^{n_z} \rightarrow \mathbb{R}$ is a continuous, 
positive definite kernel function and $K_\datasetdim$ is the resulting kernel matrix $[k(z,z^{\prime})]_{z,z^{\prime} \in Z}$. 
The posterior variance $\gppostvar^2(z)$ quantifies the epistemic uncertainty of the model arising due to the lack of data.
The kernel $\kernelfunc$ induces a reproducing kernel Hilbert space (RKHS) $\RKHS$ \cite{kanagawa2018gaussian} and by definition \eqref{eq:mean_update}, the posterior mean lies in the associated RKHS, $\mu\in \RKHS$. The space $\RKHS$ is equipped with the RKHS norm $\RKHSnorm[\cdot]$, which characterizes the regularity of functions in RKHS.

We denote $\dynGP \sim \GP(\mu, k_D)$ to represent the sampled dynamics function $g$ drawn from a GP characterized by mean $\mu$ and covariance function (kernel) $\kernelfunc_D$.
Using the kernel, we next make a standard regularity assumption on unknown dynamics $\dynGPtrue$~\cite{koller_learning-based_2018,prajapat2025safe,prajapat2022near,vakili2021scalable,abbasi2013online,beta_srinivas}: 
\begin{assumption}[Regularity~\cite{prajapat2025safe}]
    \label{assump:q_RKHS}
    \looseness -1  
    The unknown dynamic~$\dynGPtrue$ is an element of the 
    RKHS $\RKHS$ associated with the kernel $\kernelfunc$, 
    having a known bound $\Bg$ on the RKHS norm, i.e., \mbox{$\dynGPtrue \in \RKHS$ } with \mbox{$\RKHSnorm[\dynGPtrue] \leq \Bg < \infty$}. 
\end{assumption}

\mypar{Objective} Our objective is to predict a \emph{reachable set} that contains the true trajectory resulting from the unknown dynamics \eqref{eq:system_dyn} with a user-specified, arbitrary high probability \mbox{$1-\delta \in (0,1)$}. 
We achieve this by directly sampling a finite number of continuous dynamics from the posterior distribution, effectively capturing the epistemic uncertainty characterized by the GP model.
We simulate each sample and collectively build a reachable set. Ultimately, our goal is to utilize this set to design an MPC that is recursively feasible and guarantees closed-loop safety and stability of the unknown system \eqref{eq:system_dyn}.

%
\section{Finite sampling to treat epistemic uncertainty}
\label{sec:finite_sample_complexity}
This section addresses the following question: 
\begin{tcolorbox}[colframe=white!, top=2pt,left=2pt,right=2pt,bottom=2pt]
\centering
\emph{How many dynamics functions must be sampled to effectively treat the epistemic uncertainty up to an arbitrary small tolerance $\epsilon>0$?}
\end{tcolorbox}
For this, we derive a sample complexity bound in \cref{sec:sample_complexity} and then extend it to provide rate results with respect to tolerance $\epsilon$ for commonly used kernels in \cref{sec:sample_complexity_rate}.

\subsection{Sample complexity}
\label{sec:sample_complexity}
We aim to determine the number of samples $g\sim\GP(\mu,k_D)$ required to ensure that at least one is uniformly $\epsilon$-close to the unknown function. The key challenge lies in quantifying the probability of a \emph{continuous} function (of infinite dimension) being contained in an $\epsilon$-ball around a sampled function.
For this, we leverage the concept of small ball probability from the stochastic-processes literature~\cite{li2001gaussianSBP,stolz1996some,van2011information}:

\begin{definition}[\!\cite{van2011information}]  
\label{def:sbp} Let $\dynGP \sim \GP(0, k)$. The small ball probability is defined  as $\probability{ \|\dynGP\|_{\infty} < \epsilon} \eqqcolon e^{-\phi(\epsilon)}$ for any $\epsilon > 0$, where the small ball exponent $\phi(\epsilon): \R_{>0} \to \R_{>0}$.
\end{definition}
Henceforth, we refer to $\phi(\epsilon)$ as the small ball exponent. The small ball exponent depends on the choice of prior kernel $k$, and upper bounds can be established based on the kernel's properties. 
For instance, the squared-exponential kernel satisfies \mbox{$\phi(\epsilon) \leq C \left( \log ({1}/{\epsilon})\right)^{1+d}$}; in the case of $\nu$-regular Matern kernels, $\phi(\epsilon) \leq C \left(1/{\epsilon}\right)^{d/\nu}$ \cite[Lemma 3 and 6]{van2011information}. 
Furthermore, upper bounds on $\phi(\epsilon)$ can be derived using the covering number of the unit ball in the RKHS space \cite{li1999approximation,kuelbs1993metric}. Analogously with some regularity on the kernel function via the canonical metric $d^2_k(z,z') \leq | z-z'|^\gamma_2$, similar bounds can be derived, see \cite{stolz1996some}.

Although the sampled function does not belong to the same RKHS with probability one~\cite{kanagawa2018gaussian}, this is not limiting for our setting, as we are only concerned with achieving $\epsilon$-closeness in the infinity norm, as defined in \cref{def:sbp}.


The small ball probability characterizes the probability that the sampled function falls in a ball around the zero-mean function. However, we need to quantify the probability that the sample falls in a ball around the unknown function $g^\star$. To relate these probabilities, we use the following lemma from \cite{van2008reproducing}:

\begin{lemma}[{\cite[{Lemma 5.2}]{van2008reproducing}}] \label{lem:Cameron_Martin_thm} 
Consider $\dynGP\sim\mathcal{GP}(0,k)$. 
For any $h\in \RKHS$ and any Borel measurable set $C\subseteq \mathbb{B}$ with  $h\in C \leftrightarrow -h\in C$: 
\begin{align}\label{eq:SBP_shift}
    \probability{\dynGP - h \in C} \geq \expo{-\frac{1}{2}\RKHSnorm[h]} \probability{ \dynGP\in C}.
\end{align}    
\end{lemma}
Lemma~\ref{lem:Cameron_Martin_thm} characterizes how the probability decays when the function is shifted by a continuous function $h$ in the same RKHS space. In our case, this shift is from the mean function to the unknown function, i.e., $\dynGPtrue - \mu$. Since $\dynGPtrue$ is unknown, the following lemma establishes an upper bound on this quantity.




\begin{lemma} \label{lem:C_D_norm_bound}
Let \cref{assump:q_RKHS} hold. 
Then, it holds that $\probability{\RKHSdatanorm[\dynGPtrue -\mu]^2/2 \leq C_D}\geq 1-\delta/2$ with the known constant 
\begin{multline}
\label{eq:def_C_D}
C_D \coloneqq \big( \Bg^2 - \RKHSnorm[\mu]^2 + 2\sum\nolimits_{i=1}^{D} |\alpha_i| \sqrt{\betadata[D]}\sigma(z_i) \\ + \lambda^{-2}\betadata[D] \sum\nolimits_{i=1}^{D} \sigma(z_i)^2 \big)/ 2 >0.
\end{multline}
\end{lemma}

The proof can be found in \cref{apx:aux_lemma}. The above lemma utilizes high-probability confidence bounds of GPs~\cite{abbasi2013online}, where $\betadata[D]$ is the corresponding scaling factor 
(cf. \cref{thm: beta} in \cref{apx:aux_lemma}).
Secondly, this result relates the RKHS norm of the prior and posterior kernel using  $\RKHSdatanorm[\dynGP]^2 = \RKHSnorm[\dynGP]^2 + \lambda^{-2} \sum_{i=1}^{D} \dynGP(z_i)^2 $, $\forall \dynGP \in \RKHS$, see \cite[Appendix B]{beta_srinivas}.
Lastly, note that $C_D$ is a data-dependent constant and can be computed exactly.

Next, we present our main result, which bounds the number of continuous samples from the GP posterior required to ensure that at least one is uniformly $\epsilon$-close to the unknown function.

\begin{theorem} \label{thm:sample_complexity} Let \cref{assump:q_RKHS} hold and consider any  $\epsilon>0$. Let $\numsamples \in \Nat$ be a total number of GP samples $\dynGP^\n \sim GP(\mu, k_D)$ with 
\begin{align}
    \numsamples \geq \numsamples_{\epsilon} \coloneqq \frac{\log\left(\delta/2\right)}{\log \left(1 - \expo{- \left(C_D +\phi(\epsilon)\right)} \right) }\label{eqn:sample_complexity}
\end{align}
and $C_D$ as defined in \cref{lem:C_D_norm_bound}.
Then $\exists \n \in \Intrange{1}{\numsamples}$ such that 
$ \inftynorm[\dynGP^\n - \dynGPtrue] < \epsilon$ with probability at least $1-\delta$.
\end{theorem} 
\begin{proof} 
First, we bound the probability that one sample from the GP posterior is uniformly $\epsilon$ close to the true function. Later, we use this to determine the number of samples~$\numsamples$  needed to ensure at least one sample is close to the true function with probability $1-\delta$.


Suppose $\RKHSdatanorm[\dynGPtrue-\mu]^2/2 \leq C_D$. Then, \cref{lem:Cameron_Martin_thm} implies,
\begin{align*}
        \probability{ \| \dynGP^\n - \dynGPtrue \|_{\infty} < \epsilon} &= \probability{ \|\dynGP^\n - \mu - (\dynGPtrue - \mu) \|_{\infty} < \epsilon}\\
        &\labelrel\geq{step:cameron_martin} e^{-\frac{1}{2} \RKHSdatanorm[\dynGPtrue-\mu]^2} \probability{\|\dynGP^\n - \mu \|_{\infty} < \epsilon} \\
        &\labelrel\geq{step:known_constant_c_D}  e^{- C_D} \probability{\|\dynGP^\n - \mu \|_{\infty} < \epsilon} \\
        & \labelrel\geq{step:posterior_dist_to_prior}  e^{- C_D} \probability{\|\dynGP \|_{\infty} < \epsilon} \\
        &\eqqcolon e^{-\left(C_D+\phi(\epsilon)\right)} \numberthis \label{eq:sample_prob}
\end{align*}
where $\dynGP \sim\mathcal{GP}(0, k)$ and $\phi(\epsilon) \!=\! -\log  \probability{\|\dynGP\|_{\infty} < \epsilon}$ from \cref{def:sbp}. 
In the proof above, \eqref{step:cameron_martin} follows from \cref{lem:Cameron_Martin_thm} since, 
$C = \{f\in\mathbb{B}~|~\|f \|_{\infty} <\epsilon\}$
is a symmetric set $(f\in C \leftrightarrow -f\in C)$, $(\dynGP^\n - \mu) \sim \mathcal{GP}(0, k_\datasetdim)$
and $\dynGPtrue -\mu \in \RKHS$ since $\dynGPtrue \in \RKHS$, $\mu \in \RKHS$. 
The last step \eqref{step:posterior_dist_to_prior} follows from \cref{lem:small_ball_comparison_prior_posterior_variance} in \cref{apx:aux_lemma}.

Using \cref{eq:sample_prob} we obtain that
\begin{align*}
    \probability{ \|\dynGP^\n - \dynGPtrue\|_{\infty} < \epsilon}  &\geq e^{-\left(C_D+\phi(\epsilon)\right)} \\
   \iff    \probability{ \|\dynGP^\n - \dynGPtrue\|_{\infty} \geq \epsilon}  &<  1- e^{-\left( C_D +\phi(\epsilon)\right)}.
 \end{align*}
Since the dynamics samples $\dynGP^\n \sim GP(\mu, k_D)$ are drawn independently, the probability of all samples simultaneously not being $\epsilon$-close uniformly is 
 \begin{align*}
&\!\!\!\!\!\!\probability{ \|\dynGP^\n - \dynGPtrue\|_{\infty} \!\geq \epsilon, \forall \n \in \Intrange{1}{\numsamples}} \! < \! \left(1- e^{-\left(C_D+\phi(\epsilon)\right)}\right)^{\!\!\numsamples} \\
\!\!\!\!\iff\!\! &\probability{\exists \n \! \in\! \Intrange{1}{\numsamples}\!\!:\!\|\dynGP^\n \!-\! \dynGPtrue\|_{\infty} \!\!<\! \epsilon \!} \! \geq \!1\! - \!\left(\!1\!- e^{\!-\left(C_D+\phi(\epsilon)\right)}\!\right)^{\!\!\numsamples}
\end{align*}
Given that the number of samples $\numsamples \in \Nat$ satisfies \eqref{eqn:sample_complexity}, 
we have
\begin{align}
     \probability{ \exists \n \in \Intrange{1}{\numsamples} : \|\dynGP^\n - \dynGPtrue\|_{\infty} \leq \epsilon} \geq 1 - \delta/2. \label{eq:N_to_delta_a_sample_exists}
\end{align}
Note that the $C_D$ upper bound used in step \eqref{step:known_constant_c_D} in \eqref{eq:sample_prob} holds with probability $1-\delta/2$ (cf. \cref{lem:C_D_norm_bound}). Together with~\eqref{eq:N_to_delta_a_sample_exists}, both events hold jointly with probability $1-\delta$ using Boole's inequality. 
\end{proof}
By sampling $\numsamples$ continuous functions from the GP posterior, we are guaranteed, with high probability, to have at least one sample within an $\epsilon$-ball around the unknown function. 
Note that, for any user-defined arbitrarily small tolerance $\epsilon > 0$ and arbitrarily high probability $ 1-\delta \in (0,1)$, we can compute a finite number of samples $N$ using~\eqref{eqn:sample_complexity}. 
We refine the analysis for the special case of bounded noise in \cref{sec:sample_complexity_bounded_noise}.


Note that while our sample complexity result involves sampling from a Gaussian Process — similar in spirit to Thompson sampling~\cite{thompson1933likelihood,vakili2021scalable}— we do not assume that the true dynamics follow this distribution. This contrasts with scenario-based approaches~\cite{devonport2020estimating,umlauft_scenario-based_2018-1,Campi2019Scenario,garatti2021risk,lefringhausen2024learning}, which require that the samples and the ground truth follow the same distribution. Our method assumes that the unknown dynamics lies in the associated RKHS.

\begin{remark}[Extension to vector-valued $\dynGPtrue$] \label{rem:vector_valued}
To extend the analysis to vector-valued functions $\dynGPtrue:\mathbb{R}^{n_x+n_u}\rightarrow\mathbb{R}^{n_g}$, we can model each component independently as a GP (assuming independent noise across components).
This allows us to apply \cref{lem:Cameron_Martin_thm} independently along each dimension. Thus, \cref{eq:sample_prob} results in $$\probability{ \inftynorm[\dynGP^\n_i - \dynGPtrue_i] < \epsilon_i,~\forall i\in\Intrange{1}{n_g}} \leq  e^{-\left(\sum\nolimits_{i=1}^{n_g}C_{D_i}+\phi(\epsilon_i)\right)}.$$ This bounds the probability that all components simultaneously stay within $\epsilon_i$ neighbourhoods of $\dynGPtrue_i$ 
    and consequently, the sample complexity result \eqref{eqn:sample_complexity} follows analogously with the factor $\sum_{i=1}^{n_g}C_{D_i}+\phi(\epsilon_i)$.
\end{remark}

\subsection{Sample complexity rates}
\label{sec:sample_complexity_rate}
The following corollary derives the rate at which the number of samples $N$ grows as $\epsilon \to 0$ for commonly used kernels.
\begin{corollary}[Sample complexity rates] Let \cref{assump:q_RKHS} hold 
and 
consider $\numsamples_\epsilon$ from \cref{thm:sample_complexity} and a GP input dimension of $d\in\Nat$. As $\epsilon \to 0$, the number of samples $\numsamples_\epsilon$ scales with the following rates:
\begin{itemize}
    \item For the squared exponential kernel:
    \begin{align*}
\numsamples_\epsilon &= \mathcal{O}\left(\left({1}/{\epsilon}\right)^{C\left(\log ({1}/{\epsilon})\right)^d}\right); 
    \end{align*}
    \item For $\nu$-regular Mat\'ern kernels:
    \begin{align*}
        \numsamples_\epsilon &= \mathcal{O}\left(e^{C\left({1}/{\epsilon}\right)^{d/\nu}}\right).
    \end{align*}

\end{itemize}
\end{corollary}
\begin{proof} 
Note that as $s \to 0$, it holds that $s \leq \log (1/(1-s))$. 
With $s = e^{-\left(C_D+\phi(\epsilon)\right)}$ we can bound $\numsamples_\epsilon$ as:
    \begin{align*} 
        \numsamples_\epsilon &\leq \frac{\log(2/\delta)}{e^{-\left(C_D+\phi(\epsilon)\right)}}
        = \mathcal{O}\left(e^{\phi(\epsilon)}\right).   \end{align*}
For the squared-exponential kernel, using $\phi(\epsilon) \leq C \left( \log \frac{1}{\epsilon}\right)^{1+d}$ \cite[Lemma 6]{van2011information}, we obtain
    \begin{align*}
      \numsamples_\epsilon  = \mathcal{O}\left(e^{C\left(\log ({1}/{\epsilon})\right)^{1+d}}\right) 
= \mathcal{O}\left(\left({1}/{\epsilon}\right)^{C\left(\log ({1}/{\epsilon})\right)^d}\right).
\end{align*}
For $\nu$-regular Mater\'n kernel using $\phi(\epsilon) \leq C \left(\frac{1}{\epsilon}\right)^{d/\nu}$ \cite[Lemma 3]{van2011information}, we obtain
\begin{align*}
      \numsamples_\epsilon  = \mathcal{O}\left(e^{C\left({1}/{\epsilon}\right)^{d/\nu}}\right).\quad\quad\quad\qedhere
\end{align*}
\end{proof}
For the Mat\'ern kernel, the number of samples increases exponentially with $1/\epsilon$. Similarly, for the squared-exponential kernel, the number of samples grows exponentially, however, with an exponent that only has a logarithmic dependence on $1/\epsilon$, resulting in a slower growth rate.






\section{Sampling-based reachable set}
\label{sec:reachable_set}


Using the finite number of samples drawn from the GP posterior, we now formulate a sampling-based reachable set that contains the true trajectory with high probability.
We sequentially propagate uncertainty arising from the remaining $\epsilon$-epistemic uncertainty and the aleatoric noise $w$ by assuming Lipschitz continuity of the dynamics and bounded process noise.


\begin{assumption} \label{assump:lipschitz_dyn}
    The true dynamics $\dyntrue \coloneqq \dynNom + \gpsubspace \dynGPtrue$ is $\Lipdyn$-Lipschitz continuous with respect to $\state$, i.e., 
    \begin{align} \label{eq:lipschitz}
        \euclidnorm[\dyntrue(\state, \coninput) - \dyntrue(\state^\prime, \coninput)] \leq \Lipdyn \euclidnorm[\state-\state^\prime]\quad \forall x,x'\in\mathbb{R}^n.
    \end{align}
\end{assumption}
\looseness -1 Under \cref{assump:q_RKHS}, the unknown function is Lipschitz continuous for kernels that are Lipschitz continuous, such as the squared exponential and Mat\'ern kernels ~\cite{fiedler2023lipschitz}.


\begin{assumption} \label{assump:bounded_noise}
There exists a known constant $\noisebound$ that bounds the process noise, i.e., $|\noise(\ki)| \!\leq\! \noisebound$ for all times $\ki \in \Intp$\!.
\end{assumption}
We assume bounded process noise in order to construct a bounded reachable set, see also~\cref{rem:other_propagation_methods} regarding the handling of unbounded noise.



Let $\state^\n_\ki$ denote the state according to the nominal dynamics, given by the GP sample $\dynGP^\n$ at the $k^{th}$ time step, i.e., 
\begin{align}\label{eq:sample_dyn_propagation}
    \!\!\!\state^\n_{\ki+1} = \dynNom(\state^\n_\ki, \coninput(\ki)) + \gpsubspace \!\left( \dynGP^\n(\state^\n_\ki, \coninput(\ki)) \right)\!,\!
\end{align}
for a given input sequence $\coninput(\ki), k \in \Intp$, and initial state $\state^n_0=\state(0)$. 
We construct a ball \mbox{$\ball{\epsmpc}\! = \!\{x\!\in\R^{n_x} | \|x\|\!\leq\!\epsmpc\}$} around the states  $\state^\n_\ki$ reached by the samples to account for uncertainties in the reachable set. The following theorem formally defines the reachable set and establishes that it contains the true trajectory resulting from unknown dynamics~\eqref{eq:system_dyn} with high probability. 

\begin{theorem}[Reachable set] \label{thm:PRS} Let \cref{assump:q_RKHS,assump:lipschitz_dyn,assump:bounded_noise} hold. Consider any $\epsilon>0$ and $N$ according to \cref{thm:sample_complexity}. For any initial state $\state(0)\in\mathbb{R}^n$ and any control input sequence $\bm{u}_k = [\coninput(0), \hdots, u(\ki-1)]$, define the reachable set at any time step $k\in\Intp$ as
\begin{align}
    \mathcal{R}_{\ki}(x(0), \bm{u}_k)  \coloneqq \bigcup_{\n \in \Intrange{1}{N}} \state^\n_\ki \oplus \ball{\epsmpc[\ki]}, \label{eqn:PRS_definition}
\end{align}
where $\epsmpc[\ki] \coloneqq \epsdyn L_{{\ki}}$, $\epsdyn \coloneqq \euclidnorm[\gpsubspace] (\epsilon + \noisebound)$  and $L_{{\ki}} \coloneqq  \sum_{i=0}^{\ki-1} \Lipdyn^i$. Then it holds that
\begin{align}\label{eq:reachable_set_guarantee}
       \probability{\state(i) \in \mathcal{R}_{i}(x(0), \bm{u}_i),~\forall i \in \Intrange{0}{\ki}} \geq 1- \delta.
\end{align} 
\end{theorem} 
\begin{proof}
Suppose the $\n^{th}$ dynamic sample satisfies $\inftynorm[\dynGP^\n - \dynGPtrue] \leq \epsilon$. We first bound the deviation between $\state(i)$ from $\state^\n_{i}$. For any given control input $\coninput(i)$, it holds that
\begin{align*}
&\euclidnorm[\state(i+1)-\state^{\n}_{i+1}]\\
&\qquad= \|\dynNom(\state(i), \coninput(i))+ \gpsubspace(\dynGPtrue(\state(i), \coninput(i)) + w(i)) \\ 
&\ \ \quad\qquad\qquad\qquad - \dynNom(\state^{\n}_{i}, \coninput(i)) -\gpsubspace\dynGP^{\n}(\state^{\n}_{i}, \coninput(i))\|\\
&\qquad\stackrel{\eqref{eq:lipschitz}}{\leq} \!\Lipdyn\euclidnorm[\state(i)-\state^{\n}_{i}]+\underbrace{\euclidnorm[\gpsubspace]( \inftynorm[\dynGPtrue-\dynGP^{\n}] + |w(i)|)}_{\leq \epsdyn}.
\end{align*}
Iteratively applying this with $\state(0)=\state^{\n}_{0} $ yields
\begin{align*}
&\|\state(i)-\state^{\n}_{i}\|\leq  \underbrace{\left( \sum_{j=0}^{i-1} \Lipdyn^j \right)}_{L_{{i}}} \epsdyn = \epsdyn L_{{i}} \coloneqq \epsmpc[i].
\end{align*}

This implies \mbox{$\state(i) \in \state^\n_{i} \oplus \ball{\epsmpc[i]}$} if \mbox{$\exists \n : \inftynorm[\dynGP^\n - \dynGPtrue] \leq \epsilon$}. Using \cref{thm:sample_complexity}, there exists at least one out of $N$ samples satisfying $\inftynorm[\dynGP^\n - \dynGPtrue] \leq \epsilon$ with probability at least $1-\delta$. Hence, with the same probability, the true state $\state_{i}$ lies in the union of these balls, $\mathcal{R}_{i}(x(0), \bm{u}_i)$, i.e., \cref{eq:reachable_set_guarantee} holds. 
\end{proof}

The reachable set~\eqref{eqn:PRS_definition} propagates the epistemic uncertainty through samples, thereby avoiding the conservatism inherent in common sequential propagation methods~\cite{marruedo2002ISSrobustMPC,sasfi2023robust,Houska2019robustMPCbook,koller_learning-based_2018}. 
Only the remaining $\epsdyn$ uncertainty, which captures the residual $\epsilon$-epistemic uncertainty and bounded aleatoric noise $\noisebound$ is propagated sequentially in the proposed approach. As established in \cref{thm:sample_complexity}, choosing a sufficiently large number of samples $N$ allows $\epsilon>0$ to be arbitrarily small, effectively treating the epistemic uncertainty of the unknown dynamics. 
While sequential propagation of the remaining uncertainty $\epsdyn$ causes the reachable set to grow exponentially with the horizon when the Lipschitz constant $\Lipdyn>1$, the resulting growth is significantly scaled down in contrast to sequentially propagating the full uncertainty.
In particular, for small noise $w$ and with a large number of samples $N$, the balls are small, resulting in a valid sampling-based reachable set with minimal conservatism in contrast to the sequential propagation techniques. 


\begin{remark}[Alternative propagation techniques]\label{rem:other_propagation_methods}
For simplicity, we used a simple Lipschitz continuity-based propagation similar to~\cite{marruedo2002ISSrobustMPC} to address the aleatoric and the remaining $\epsilon$-epistemic uncertainty. However, this can be naturally replaced by other sequential uncertainty propagation techniques from the non-linear robust reachability literature. These include contraction metrics~\cite{sasfi2023robust}, Taylor series~\cite{althoff2008reachability} or interval arithmetics~\cite{limon2005robust}; see 
\cite{Houska2019robustMPCbook} for an overview. 
Moreover, a state-feedback $u(k)=\kappa(x(k),v(k))$ can be employed to reduce conservatism in the sequential propagation of  the balls $\ball{\epsmpc[\ki]}$. 
In particular, a simple modification is to use a linear state-feedback controller $u=Kx+v$ and a weighted norm $\|\state\|_P \coloneqq \sqrt{\state^\top P \state}$ with a positive definite matrix $P \succ 0$, yielding the reduced Lipschitz constant $L_P$: 
\begin{align}
\!\!\!    \| \dyntrue(\state, K \state \!+\! v) \!-\! \dyntrue(\state', K \state'\! +\! v)  \|_P \leq \!\Lipdyn_P \| \state - \state' \|_{P}, \!\! \label{eq:lip_norm}
\end{align}
which is a special case of the general approach using contraction metrics~\cite{sasfi2023robust}. \\
\cref{assump:bounded_noise} on bounded noise can also be relaxed by using the covariance (proxy) of the sub-Gaussian distribution \cite{ao2025stochastic} and recent nonlinear stochastic reachability techniques~\cite{kohler2024predictive,liu2025safety}.
\end{remark}



\begin{remark}~\label{rem:vector_valued_noisebound}
    In the case of vector-valued $\dynGPtrue$, \cref{thm:PRS} holds analogously with $\varepsilon$ such that  $$\sum_{i=1}^{n_g} \euclidnorm[{\gpsubspace}_{,i}]( \inftynorm[\dynGPtrue_i -\dynGP^{\n}_i] + |w_i(k)|) \leq \varepsilon,$$ using also the bound from Remark~\ref{rem:vector_valued}.
\end{remark}

Based on this probabilistic reachable set \eqref{eqn:PRS_definition}, we can directly formulate an optimal control problem that ensures constraint satisfaction with probability at least $1-\delta$ for a finite horizon. 
In the next section, we present an MPC formulation that yields closed-loop guarantees for receding-horizon implementation.

\section{Model predictive control}
\label{sec:MPC}
This section develops a control strategy that minimizes a stage cost \mbox{$\cost: \mathbb{R}^{n_\state} \times \mathbb{R}^{n_\coninput}\rightarrow \mathbb{R}$}, while ensuring that the closed-loop system satisfies the joint chance constraints
\begin{align}
    \probability{ \state(\ki) \in \stateSpace, \coninput(\ki) \in \inputSpace,~\forall \ki \in \Intp} \geq 1-\delta, 
    \label{eq:constraints_path}
\end{align} 
despite the presence of both aleatoric and epistemic uncertainty of $\dynGPtrue$. 
A key challenge in ensuring this during closed-loop operation is maintaining recursive feasibility, due to non-sequential uncertainty propagation by the sampling-based reachable set. Unlike robust MPC frameworks that rely on sequential uncertainty propagation, such tools~\cite{marruedo2002ISSrobustMPC,sasfi2023robust,Houska2019robustMPCbook} cannot be directly applied here.
To address this, we propose a strategy to retain only a subset of samples over time while ensuring that the sample that is $\epsilon$-close to the unknown dynamics $g^\star$ is retained.

Building on the reachable set from \cref{sec:reachable_set}, we first formulate a robust sampling-based GP-MPC problem in \cref{sec:MPCformulation}. In \cref{sec:update_dyn_set}, we present a sample selection mechanism that is crucial for maintaining recursive feasibility. 
Finally, we provide high-probability safety guarantees for the unknown system using the finite set of samples in \cref{sec:rec_feas_safety_guarantees} and later show the stability and performance in \cref{sec:stability}.


\subsection{MPC formulation}
\label{sec:MPCformulation}
The proposed MPC problem at time~$k \in \Intp$ is given by
\begin{subequations} \label{eq:SafeGPMPCinf}
    \begin{align}
        \underset{\bm{\state}^\n, \bm{u}}{\min} \quad & 
        \sum_{\n \in \N_{\ki}} 
        \Bigg[ 
        \sum_{i=0}^{\horizon-1} \cost(\state^\n_{i|\ki}, \coninput_{i|\ki} )  + \terminalcost{\state^\n_{\horizon|\ki}} \Bigg] 
        \label{eq:SafeGPMPCcost}
        \\
        \mathrm{s.t.} \quad &\forall i \in \Intrange{0}{\horizon-1}, 
        \forall \n \in \N_{\ki}, 
        \label{eq:SafeGPMPCinf_forall} \\
        & \state^\n_{0|\ki} = \state(k), \label{eqn:initialization}\\
        & \state_{i+1|\ki}^\n = \dynNom(\state^\n_{i|\ki}, \coninput_{i|\ki}) + \gpsubspace \dynGP^\n(\state^\n_{i|\ki},\coninput_{i|\ki}), \label{eqn:pred_dyn}\\
        & \state^\n_{i|\ki} \in \stateSpace \ominus  \ball{\epstightening[i]}, ~\coninput_{i|\ki} \in \inputSpace \label{eqn:state_input_constraint}, \\
        & \state^\n_{\horizon|\ki} \in \terminalset \ominus \ball{c_{\horizon-1}}. \label{eqn:terminal_constraint}
    \end{align}
\end{subequations}

\looseness -1 
This non-linear program is solved at each time step $\ki$ using a subset of sampled dynamics $\N_{\ki}  \subseteq \Intrange{1}{\numsamples}$. 
We optimize an open-loop (shared) input sequence $\bm{\coninput} = (\coninput_0, \ldots, \coninput_{\horizon - 1})$ to predict a state sequence \mbox{$\bm{\state}^\n = (\state^\n_0, \ldots, \state^\n_\horizon)$} for each dynamics sample $\n \in \N_{\ki}$. 
\cref{eqn:state_input_constraint} ensures that all the predicted sequences satisfy the state and input constraints given by a tightened version of \eqref{eq:constraints_path}. 
The tightening $\epstightening[i]$ in \cref{eqn:state_input_constraint} is defined as $\epstightening[i] \coloneqq \sum_{j=0}^{i-1} c_j$ where $c_j$ accounts for the remaining $\epsilon$-epistemic uncertainty and the aleatoric noise; see \cref{lem:tightening} below for its derivation.
The constraint in \cref{eqn:initialization} initializes all the trajectories to the system state at time $\ki$, denoted as $\state(\ki)$, and \cref{eqn:terminal_constraint} ensures that all the predicted trajectories reach the tightened version of the terminal set $\terminalset \subseteq\R^{n_x}$
at the end of the horizon. 
The cost in~\cref{eq:SafeGPMPCcost} is the empirical average of finite-horizon and terminal cost over the dynamics contained in the set $\N_\ki$ (modulo scaling constant), with stage cost $\cost$
and terminal cost \mbox{$\terminalcostNA : \mathbb{R}^{n_\state} \rightarrow \mathbb{R}$}. 
We assume $\ell$, $\terminalcostNA$, $\dynGP^\n$, $f$ are continuous and the sets $\stateSpace, \inputSpace, \terminalset$ are compact, which implies a minimizer of Problem~\eqref{eq:SafeGPMPCinf} exists \cite[Proposition 2.4]{rawlingsModelPredictiveControl2020}.

\cref{alg:rhc_update_Set} outlines the steps in the receding-horizon implementation of Problem~\eqref{eq:SafeGPMPCinf}. Next, we discuss 
how the dynamics set $\mathcal{N}_k$ is updated by removing falsified samples.

\begin{algorithm}[t]
\caption{Receding horizon Sampling GP-MPC}
\begin{algorithmic}[1]
\State Pick \mbox{$1\!-\!\delta \!\in\! (0,1)$}, $\epsilon\!>\!0$, $D$ data points, compute $\numsamples$\eqref{eqn:sample_complexity}
\State Build $\N_0$ by sampling $\numsamples$ dynamics, $\dynGP^\n \sim \GP(\mu, \kernelfunc_D)$ 
\State Design terminal ingredients $\terminalset, \terminalcostNA$ and tightenings $\epstightening$
\For{time $\ki = 0,1, \hdots $} \label{alg: termination-condi}
\State $\bf{u}^\star \xleftarrow{}$ Solve Problem \eqref{eq:SafeGPMPCinf} with state $x(\ki)$
\State Apply the first input $u(\ki)=u^\star_{0|\ki}$.
\State Update dynamics set $\N_{\ki+1}$ as per \cref{eqn:adaptive_dyn_set}.
\EndFor
\end{algorithmic}
\label{alg:rhc_update_Set}
\end{algorithm}

\subsection{Updating the finite set of dynamics samples}
\label{sec:update_dyn_set}

In Problem~\eqref{eq:SafeGPMPCinf}, $\N_{\ki}, ~ \ki\in \Intp$ is a finite set of dynamics sampled from the GP distribution, and is monotonically shrinking with time $k$ as follows 
\begin{align}\label{eqn:adaptive_dyn_set}
\!\!\!\N_{\ki+1} \!=\! \{ \n \!\in\! \N_{\ki}|\,\|x^\n_{i+1|\ki} \!\!-\! x^\n_{i|\ki+1}\!\|\! \leq\! c_i,\! \forall i \!\in\! \Intrange{0}{\horizon-1} \}\!\! 
\end{align}
with $\N_{0} = \Intrange{1}{N}$ denoting the initial set of all sampled dynamics $\dynGP^\n \sim GP(\mu, k_D)$. 
In \eqref{eqn:adaptive_dyn_set}, $x^\n_{i|\ki+1}$ is the candidate state sequence with initialization $x^\n_{0|\ki+1}=x(k+1)$ and propagated using control input 
\begin{align}\label{eq:candidate_input_sequence}
    \coninput_{i|\ki+1} = \coninput^\star_{i+1|\ki},~~\forall i \in \Intrange{0}{\horizon-2}
\end{align}
with $\bm{u}^\star$ optimized from time $k$. 
The constant $c_i$'s are determined such that the $\epsilon$-close dynamic sample $g^n$ remains in $\N_k, \forall k$. Removing the other dynamics from the set $\mathcal{N}_k$ reduces conservatism and is also required for the recursive feasibility proof. 
The following lemma determines the constants $c_i$, 

\begin{lemma} \label{lem:tightening} 
Let \cref{assump:q_RKHS,assump:lipschitz_dyn,assump:bounded_noise} hold. Suppose that \mbox{$\inftynorm[{\dynGP^\n -\dynGPtrue}] \leq \epsilon$} and Problem~\eqref{eq:SafeGPMPCinf} is feasible at time $k$. 
Consider the candidate state sequence $x^\n_{i|\ki+1}$ corresponding to the candidate input sequence~\eqref{eq:candidate_input_sequence} and dynamics~\eqref{eqn:pred_dyn}.
Then, it holds that  
$\|x^\n_{i|\ki+1} - x^\n_{i+1|\ki} \| \leq c_{i}, \forall i \in \Intrange{0}{\horizon-1}$, 
 where, 
 \begin{align}
      c_{i} \coloneqq \Lipdyn^{i} \epsdyn + 2 \|\gpsubspace\| \epsilon \sum_{j=0}^{i-1} \Lipdyn^j \label{eq:ci_tightenings}
 \end{align}
\end{lemma}
\begin{proof}
We first bound the one-step prediction error: 
\begin{align*}
&\euclidnorm[x^\n_{0|\ki+1} - x^\n_{1|\ki}]\\
&\qquad\quad= \|\dynNom(x(\ki), \coninput^\star_{0|\ki}) +  \gpsubspace (\dynGPtrue(x(\ki), \coninput^\star_{0|\ki}) + w(k))\\
     & \qquad\qquad\qquad\quad\quad - \dynNom(x(\ki), \coninput^\star_{0|\ki}) - \gpsubspace \dynGP^\n(x(\ki), \coninput^\star_{0|\ki})\| \\
&\qquad\quad\leq\euclidnorm[\gpsubspace]( \inftynorm[\dynGPtrue-\dynGP^{\n}] + \noisebound) = \epsdyn. \numberthis \label{eq:init_var_eps_condition}
\end{align*}
Similarly, we consider the subsequent prediction errors
\begin{align*}
    \|&x^\n_{i|\ki+1} - x^\n_{i+1|\ki} \| \\
    &~~= \|\dynNom(x^\n_{i-1|\ki+1}, \coninput^\star_{i|\ki}) +  \gpsubspace \dynGP^\n(x^\n_{i-1|\ki+1}, \coninput^\star_{i|\ki}) \\
    &\qquad\qquad\qquad\quad\quad - \dynNom(x^\n_{i|\ki}, \coninput^\star_{i|\ki}) - \gpsubspace \dynGP^\n(x^\n_{i|\ki}, \coninput^\star_{i|\ki})\| \\
    &~~\labelrel\leq{step:eps_close_dyn} \|\dynNom(x^\n_{i-1|\ki+1}, \coninput^\star_{i|\ki}) +  \gpsubspace \dynGPtrue(x^\n_{i-1|\ki+1}, \coninput^\star_{i|\ki}) \\
    &\qquad\quad \ \  \ - \dynNom(x^\n_{i|\ki}, \coninput^\star_{i|\ki}) - \gpsubspace \dynGPtrue(x^\n_{i|\ki}, \coninput^\star_{i|\ki})\| + 2 \|\gpsubspace\| \epsilon\\
    &~~\labelrel\leq{step:lipschitz} \Lipdyn \| x^\n_{i-1|\ki+1} - x^\n_{i|\ki} \| + 2 \|\gpsubspace\| \epsilon\\
    \!\!\!&\quad\vdots\\
    &~~\leq \Lipdyn^{i} \| x^\n_{0|\ki+1} - x^\n_{1|\ki} \| + \Lipdyn^{i-1} 2 \|\gpsubspace\| \epsilon + \hdots + 2 \|\gpsubspace\| \epsilon \\ 
    &~~\stackrel{\eqref{eq:init_var_eps_condition}}{\leq} \Lipdyn^{i} \epsdyn + 2 \|\gpsubspace\| \epsilon \sum_{j=0}^{i-1} \Lipdyn^j   \eqqcolon c_{i}. 
\end{align*} 
In the above proof, \eqref{step:eps_close_dyn} follows since $\inftynorm[{\dynGP^\n -\dynGPtrue}] \leq \epsilon$ and \eqref{step:lipschitz} using Lipschitz continuity of the dynamics (\cref{assump:lipschitz_dyn}).
\end{proof}

\begin{remark} Suppose that $\dynNom + \gpsubspace \dynGP^\n$ is $\Lipdyn_g$-Lipschitz continuous for all $\n \in \N_0$. Then, \cref{lem:tightening} remains valid with $c_{i} \coloneqq \Lipdyn^{i}_g \epsdyn$.  This simpler expression eliminates the additional summation term in \eqref{eq:ci_tightenings}. Moreover, depending on the regularity of the kernel, the GP samples are Lipschitz continuous \cite{da2023sample,lederer2019uniform}.
\end{remark}

The following corollary uses \cref{thm:sample_complexity} and the sufficient condition of \cref{lem:tightening} to formally establish that updating the dynamics set \eqref{eqn:adaptive_dyn_set} retains the $\epsilon$-close sample.
\begin{corollary} \label{coro:adaptive_set_high_prob}
Let \cref{assump:q_RKHS,assump:lipschitz_dyn,assump:bounded_noise} hold. Consider any $\epsilon>0$ and $\N_0 = \Intrange{1}{N}$ with $N$ as per \cref{thm:sample_complexity}. Then it holds that
\begin{align*}
    \probability{\forall k \in \Intp,~\exists \n \in \N_{\ki} : \|\dynGP^\n - \dynGPtrue\|_{\infty} < \epsilon} \geq 1 - \delta.
\end{align*}    
\end{corollary} 
\cref{thm:sample_complexity} ensures $\exists \n\in\N_0$ such that $\inftynorm[{\dynGP^\n -\dynGPtrue}] < \epsilon$. 
Then by \cref{lem:tightening} the same sample $\n$ can deviate at most $c_{i}$ ensuring $\|x^\n_{i|\ki+1} - x^\n_{i+1|\ki} \| \leq c_{i}, \forall i \in \Intrange{0}{\horizon-1}, \ki \in \Intp$ which is the same rule as used to update the dynamics set $\N_\ki$ \eqref{eqn:adaptive_dyn_set}. This proves \cref{coro:adaptive_set_high_prob} and ensures that the removed samples are not $\epsilon$-close.



\subsection{Closed-loop safety guarantees}
\label{sec:rec_feas_safety_guarantees}
Next, we investigate the theoretical guarantees of the proposed robust GP-MPC problem~\eqref{eq:SafeGPMPCinf}. In order to address recursive feasibility, we make the following standard assumption on the terminal invariant set.
\begin{assumption}[Robust positive invariant set] \label{assum:safeset}
The terminal set $\terminalset$ is a robust positive invariant set with the input $\terminalinput\in\inputSpace$, i.e., $\forall \n \in \N_0$, $\state \in \terminalset$: $\state \in \stateSpace \ominus \ball{\epstightening[\horizon-1]}$
and $\dynNom(\state, \terminalinput) + \gpsubspace \dynGP^\n(\state, \terminalinput) \in \terminalset \ominus \ball{c_{\horizon-1}}$.
\end{assumption}


Similar to standard MPC designs~\cite[Sec.~2.5.5]{rawlingsModelPredictiveControl2020}, this condition can be satisfied by constructing $\terminalset$ based on a local Lyapunov function, see also Section~\ref{sec:stability}.
The following theorem guarantees recursive feasibility and closed-loop 
constraint satisfaction by applying the optimal control sequence
obtained by solving problem~\eqref{eq:SafeGPMPCinf}. 
\begin{theorem} \label{thm:safety_recursive_feasibility}
Let \cref{assump:q_RKHS,assump:lipschitz_dyn,assum:safeset,assump:bounded_noise} hold. 
Consider any $\epsilon>0$ and let $\numsamples$ be the corresponding number of samples as defined in \cref{thm:sample_complexity}. Suppose that Problem~\eqref{eq:SafeGPMPCinf} is feasible at $k=0$. 
Then, with probability $1-\delta$, Problem~\eqref{eq:SafeGPMPCinf} is feasible for all $k\in\mathbb{N}$ for the closed-loop system resulting from Algorithm~\ref{alg:rhc_update_Set}. 
This further implies that the (high-probability) state and input constraints~\eqref{eq:constraints_path} are satisfied. 
\end{theorem}
\begin{proof}
Suppose that the set $\N_{\ki}$ is non-empty for all $\ki\in\Intp$, which holds with probability at least $1-\delta$ using \cref{coro:adaptive_set_high_prob}.
We show recursive feasibility using a standard proof with a candidate solution~\cite{rawlingsModelPredictiveControl2020}: 
At time $\ki + 1$, we consider the candidate input sequence
defined by \eqref{eq:candidate_input_sequence} with the terminal input $\coninput_{\horizon-1|\ki+1} = u_{\mathrm{f}}$,
where $u_{\mathrm{f}}$ ensures positive invariance of the terminal set constraint (\cref{assum:safeset}). 

We first show that the candidate input is a feasible solution to Problem~\eqref{eq:SafeGPMPCinf} at timestep $k+1$. The input constraints are trivially satisfied, i.e., $\coninput_{i|\ki+1} \in \inputSpace, \forall i \in \Intrange{1}{\horizon-1}$, due to the shifted control sequence, whose appended last input satisfies $\terminalinput\in \inputSpace$ due to \cref{assum:safeset}. 

For state constraints, \eqref{eqn:state_input_constraint} ensures \mbox{$x^\n_{i+1|\ki} \in \X \ominus \ball{\epstightening[i+1]}$,} 
$\forall \n \in \N_{\ki}$. 
The set $\N_{\ki+1}$ ensures $\|x^\n_{i|\ki+1} - x^\n_{i+1|\ki} \| \leq c_{i}, \forall i \in \Intrange{0}{\horizon-2}, \forall \n \in \N_{\ki+1}$, which implies
\begin{align}
x^\n_{i|\ki+1} \in \X \ominus \ball{\epstightening[i+1]} \oplus \ball{c_{i}} 
\subseteq  \X \ominus \ball{\epstightening[i]}. \label{eqn:state_constraint_rec_feas}
\end{align}
Analogously,~\eqref{eqn:terminal_constraint} at time $\ki$ ensures that $x^\n_{\horizon|\ki} \in \terminalset\ominus\ball{c_{\horizon-1}}$, 
which,
using the dynamics set $\N_{\ki+1}$~\eqref{eqn:adaptive_dyn_set} at \mbox{$i = \horizon-1$}, 
implies that
$x^\n_{\horizon - 1|\ki+1} \in \terminalset \ominus \ball{c_{\horizon-1}} \oplus \ball{c_{\horizon-1}} \subseteq \terminalset$.
Since $x^\n_{\horizon - 1|\ki+1}\in \terminalset$, \cref{assum:safeset} ensures the existence of a terminal input $\terminalinput$ such that $x^\n_{\horizon|\ki+1} \in \terminalset\ominus \ball{c_{\horizon-1}}$ which satisfies terminal constraint~\eqref{eqn:terminal_constraint} at $k+1$. 

Thus, in Problem~\eqref{eq:SafeGPMPCinf}, the state and input constraints are satisfied by all sampled trajectories, ensuring recursive feasibility. 
Additionally, recursive feasibility ensures $\forall \ki \in \Intp, \state(\ki) =\state_{0|k}^\n \in \stateSpace, \coninput(\ki) = u_{0|\ki} \in \inputSpace$ using ~\eqref{eqn:state_input_constraint} implying constraint satisfaction for the unknown system~\eqref{eq:system_dyn}. 
Since we assumed that the sets $\N_{\ki+1}$ are non-empty, which holds (jointly) with probability at least $1-\delta$ (\cref{coro:adaptive_set_high_prob}), all closed-loop properties are valid with probability at least $1-\delta$.
\end{proof}






Thus, for any arbitrary small tolerance $\epsilon>0$ and arbitrary high probability $1-\delta \in (0,1)$, the proposed robust sampling-based GP-MPC \eqref{eq:SafeGPMPCinf} ensures safety of the closed-loop unknown system with a finite number of dynamics samples.
Note that this methodology of rejecting falsified samples could also be interesting to ensure recursive feasibility for sampling-based approaches with finite-dimensional parameters~\cite{lew2022simple,lucia_multi-stage_2013}.

\subsection{Performance and stability}
\label{sec:stability}

In the following, we analyze the closed-loop performance and stability of the proposed MPC implementation.
Let the optimal cost of Problem~\eqref{eq:SafeGPMPCinf} at time $k$ be denoted by $J^\star(\state(\ki), \N_k)$. In order to prove stability, we make the following standard assumptions on the stage cost and the terminal cost in the terminal invariant set.

\begin{assumption}[Lipschitz cost] \label{assump:lipschitz_cost}
    The stage cost $\stagecost{\cdot}{\cdot}$ and the terminal cost $\terminalcost{\cdot}$ are $\Lipstagecost$- and $\Liptermcost$-Lipschitz continuous in $\state$.
\end{assumption}

\begin{assumption}[Terminal cost] \label{assump:terminal_cost_lyap} 
For all  $\state \in \terminalset$ and $\n \in \N_0$, the terminal cost satisfies 
   \begin{align*}
      \terminalcost{\dynNom(\state,\terminalinput) + \gpsubspace \dynGP^\n(\state,\terminalinput)} - \terminalcost{\state} \leq \!-\stagecost{\state}{\terminalinput} + \stagecostNA_s.
   \end{align*}
\end{assumption}


Assumption~\ref{assump:terminal_cost_lyap} is typically satisfied by designing a terminal set and a terminal cost around the equilibrium point~\cite{rawlingsModelPredictiveControl2020} for all dynamic samples. For example, if the origin is an equilibrium, 
i.e., $\forall \n \in \Intrange{1}{\numsamples}$, $\dynNom(0,0) + \gpsubspace \dynGP^\n(0,0)=0$, then a standard design of $V_f$~\cite{amritEconomicOptimizationUsing2011} ensures that~\cref{assump:terminal_cost_lyap} is satisfied with the equilibrium cost $\stagecostNA_s = \stagecost{0}{0}$.
This design requires that linearized dynamics around the equilibrium point are open-loop stable with a common Lyapunov function. 
The open-loop stability requirement can be relaxed by incorporating a state feedback in the predictions in~\eqref{eq:SafeGPMPCinf} as discussed in~\cref {rem:other_propagation_methods}.




Since the domain is bounded and the cost function is continuous, the cost is also bounded. 
Therefore, without loss of generality, we shift the cost such that $\stagecost{x}{u} \geq 0$ and $\terminalcost{x} \geq 0$ for all $(\state,\coninput) \in \X\times\inputSpace$.

The following theorem presents a performance bound and, with additional positive definite cost assumptions, guarantees stability for the closed-loop system by applying the optimal control sequence obtained by solving the problem~\eqref{eq:SafeGPMPCinf} in receding-horizon fashion. 
\begin{theorem}[Stability] 
\label{thm:stability}
Let \cref{assump:q_RKHS,assump:lipschitz_dyn,assump:bounded_noise,assum:safeset,assump:lipschitz_cost,assump:terminal_cost_lyap} hold and suppose that Problem~\eqref{eq:SafeGPMPCinf} is feasible at $k=0$. 
Then, with probability at least $1-\delta$, the closed-loop system resulting from Algorithm~\ref{alg:rhc_update_Set} ensures:
\begin{itemize}
        \item Asymptotic average performance bound: 
        \begin{align}
            \limsup_{T \to \infty} \frac{1}{T} \sum_{\ki=0}^{T-1} \stagecost{\state(\ki)}{\coninput(\ki)} \leq \Ljc + \stagecostNA_s \label{eqn:cost_convergence}
        \end{align}
where $\Ljc = K_1 \noisebound + K_2 \epsilon$ with constants $K_1, K_2>0$.

        \item Practical asymptotic stability:
        Suppose additionally that $\stagecost{0}{0}=0$, $\terminalcost{(0)}=0$, $\stagecost{\state}{\coninput} \geq \alpha_1(\|\state\|)$ and $  \terminalcost{\|x\|} \leq \alpha_2(\|x\|)$ with $\alpha_1, \alpha_2\in \Kinf$. Then, there exist $\beta\in\KL,\gamma_1,\gamma_2\in\mathcal{K}$:
        \begin{align}\label{eq:practical_stability}
            \| \state(\ki) \| \leq \max\{\beta(\| \state_0 \|, k ), \gamma_1(\noisebound) + \gamma_2(\epsilon)\}.
        \end{align}
    \end{itemize}
\end{theorem}
The proof is provided in \cref{apx:stability}. The constant $\Ljc$ depends linearly on the residual $\epsilon$-epsitemic uncertainty and the process noise bound $\noisebound$, and it decreases as either the process noise diminishes or $\epsilon$ shrinks with an increasing number of samples. 
Consequently, \eqref{eqn:cost_convergence} implies that,
under small process noise and residual $\epsilon$ uncertainty, the derived bound on the closed-loop cost approaches the cost at the equilibrium point $\stagecostNA_s$. 
Furthermore, \cref{eq:practical_stability} ensures that the system is practically asymptotically stable (cf.~\cite[Def.~4.1]{faulwasser2018economic}), i.e., we have asymptotic stability modulo a residual term that vanishes as the process noise  $\bar{w}$ and the remaining epistemic uncertainty $\epsilon$ approach zero. 

\section{Simulations}
\label{sec:simulation}

\begin{figure*}[t]
\begin{subfigure}[b]{0.48\textwidth}
    \includegraphics[width=1.0\textwidth]{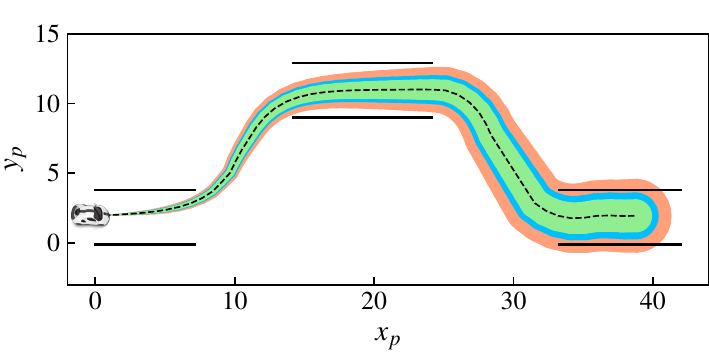}
       \caption{Finite samples based GP-MPC (ours)}
    \label{fig:samplingMPC_car}
\end{subfigure}\hfill
\begin{subfigure}[b]{0.48\textwidth}
    \includegraphics[width=1.0\textwidth]{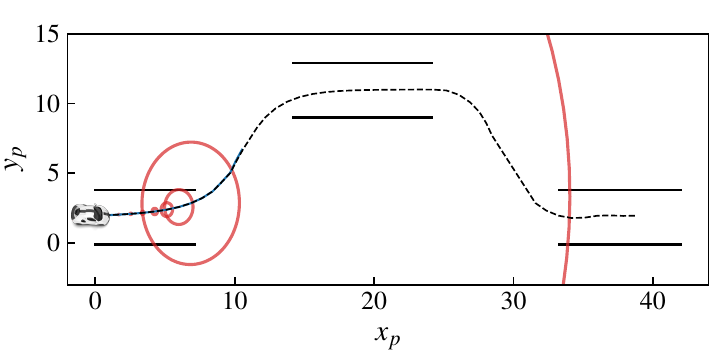}
    \caption{Robust tube-based GP-MPC~\cite{koller_learning-based_2018}}
    \label{fig:RobustMPC_car}
\end{subfigure}
\caption{
\looseness -1 Comparison of uncertainty propagation for a given input sequence $\bm{u}$ 
in the car example. The dashed line represents the true trajectory of the unknown system, navigating while satisfying the constraints (solid parallel black lines).
On the left, \cref{fig:samplingMPC_car} shows sampling-based reachable sets using $N = 2\times 10^{2}$ (red), $2\times 10^{4}$ (blue) and $2\times 10^{7}$ (green) samples. On the right, \cref{fig:RobustMPC_car} shows the reachable set over-approximation (red ellipsoids) obtained by the robust GP-MPC approach~\cite{koller_learning-based_2018}, which grows exponentially in a small horizon of 14. We observe that a tighter approximation of the reachable set achieved via the sampling-based method enables the car to navigate while satisfying the track constraints in a lane-changing maneuver, given enough samples.}\label{fig:uncertainity_car}
\end{figure*}

We demonstrate the proposed algorithm using two numerical examples. First, we show the effectiveness of the reachable set resulting from the proposed sampling-based approach (\cref{sec:reachable_set}) by comparing it to a sequential propagation technique \cite{koller_learning-based_2018} in a car environment. Second, we demonstrate closed-loop safety and recursive feasibility of sampling GP-MPC (\cref{sec:MPC}) using pendulum dynamics.

\subsection{Implementation details} \label{sec:simulation_implementaion}
In both simulations, we compute tightenings using the $P$-weighted norm Lipschitz constant as described in \cref{rem:other_propagation_methods}, which ensures that the reachable sets are not overly conservative.
%
The linear feedback $K$ and matrix $P$ are jointly optimized through linear matrix inequalities involving the Jacobian of the sampled dynamics on the constraint set, similar to~\cite{sasfi2023robust}. 
 
We used a squared-exponential kernel to model unknown dynamics in both examples. To compute the number of samples \eqref{eqn:sample_complexity}, we empirically estimate small ball probability $\phi(\epsilon)$ by evaluating the number of samples that fall within an $\epsilon$ ball around the mean using the GP posterior (\eqref{step:cameron_martin} in \cref{eq:sample_prob}). 
The constant $C_D$ is computed as per \cref{coro:sample_complexity_bounded_noise} with 
bounded noise $\noisebound$.

Note that implementing the MPC formulation \eqref{eq:SafeGPMPCinf} and the reachability analysis (\cref{thm:PRS}) requires the sampling of continuous functions \mbox{$\dynGP \sim \GP(\mu,\kernelfunc_D)$}, which cannot be implemented directly due to the infinite-dimensional feature space of GP's. 
However, equivalent trajectories can be generated by forward sampling and re-conditioning the GP, see~\cite{umlauft_scenario-based_2018-1}. 
Similarly, such forward-sampling techniques can be integrated within a sequential quadratic programming framework~\cite{prajapat2024towards} to solve the MPC problem. 
For the following MPC implementation, the open source software implementation of~\cite{prajapat2024towards} has been adapted and extended. 
The implementation is based on \texttt{Python} interfaces of \texttt{acados}~\cite{verschueren_acadosmodular_2022} and~\texttt{CasADi}~\cite{andersson_casadi_2019}; for efficient GP sampling, we employ~\texttt{GPyTorch}~\cite{gardner_gpytorch_2018}.
The source code is available at \url{https://github.com/manish-pra/sampling-gpmpc}.
All simulation parameters can be found in the params file within the repository.

\subsection{Uncertainty propagation using car dynamics}
We model the car using a kinematic bicycle model
\begin{align}
    \!\!\begin{bmatrix}
        \!\xpos(\ki\!+\!1)\!\!\\ \!\ypos(\ki\!+\!1)\!\\ \!\theta(\ki\!+\!1)\!\\ \!v(\ki\!+\!1)\!
    \end{bmatrix} \!\! &= \!\!\underbrace{\begin{bmatrix}
        \xpos(\ki) \\  \ypos(\ki) \\ \theta(\ki) \\ \!v(\ki) \!+ \!a(\ki) \Delta\!\!
    \end{bmatrix}}_{\eqqcolon \dynNom(\state,\coninput)} \!\! + \!\! 
    \underbrace{
        \begin{bmatrix} v(\ki) I_{3 \times 3} \\
        0_{1 \times 3} 
    \end{bmatrix}}_{\eqqcolon \gpsubspace}
    \!\! \underbrace{\begin{bmatrix}
    \! \cos(\theta(\ki) \!+\! \zeta_k) \Delta \!  \\ 
    \! \sin(\theta(\ki) \!+ \!\zeta_k) \Delta \! \\
    \! \sin(\zeta_k) l_r^{-1} \Delta \! 
\end{bmatrix}}_{\eqqcolon \dynGP(\state, \coninput)} 
\nonumber
\end{align}
where $\zeta_k = \tan^{-1}\left(\frac{l_r}{l_f + l_r} \tan(\delta(k))\right)$ is the slip angle~$[\si{rad}]$. The state vector $\state = [\xpos, \ypos, \theta, v]^\top$ represents the car's position $[\si{m}]$ in 2D Cartesian coordinates $[\xpos, \ypos]^\top$, its absolute heading angle $\theta$ $[\si{rad}]$, and its longitudinal velocity $v$ $[\si{m/s}]$. The control input $\coninput = [\delta, a]^\top$ consists of the steering angle $\delta~[\si{rad}]$ and the linear acceleration $a~[\si{m/s^2}]$. 
The distances between the car's center of gravity to the front and the rear wheel
are denoted by \mbox{$l_f = 1.105 [\si{m}]$} and \mbox{$l_r = 1.738 [\si{m}]$} respectively. The known part $\dynNom(\state, \coninput)$ is given by simple integrator dynamics. The unknown part, \mbox{$\dynGP: \R^2 \to \R^3$}, \mbox{$\dynGP(\state, \coninput) = \dynGP(\theta, \delta)$}, is a nonlinear function of heading and steering angle. 
Here, the projection matrix $\gpsubspace$ is scaled with velocity $v(k)$, which is a time-varying quantity known exactly since there is no uncertainty in the velocity component of the dynamics. The GP is trained with 45 prior data points (noise-free) on an equally spaced $5 \times 9$ mesh grid of the set $(\theta, \delta) = \{ [-1,1] \times [-0.6,0.6] \}$. 
We simulate reachable sets using a control sequence obtained by optimizing the trajectory of the true dynamics for an evasive maneuver test (\!\!\!\cite{kohler2019nonlinear}, compare ISO norm 3888-2 \cite{cars2011test}) as shown in \cref{fig:uncertainity_car}. 
The control sequence is optimized 
for time horizon $\horizon = 51$ with sampling time $\Delta = 60 [\si{ms}]$.
The reachable set is computed with the noise bound $\noisebound = 10^{-6}$.

\begin{figure}
    \centering
    \includegraphics[width=0.9\linewidth]{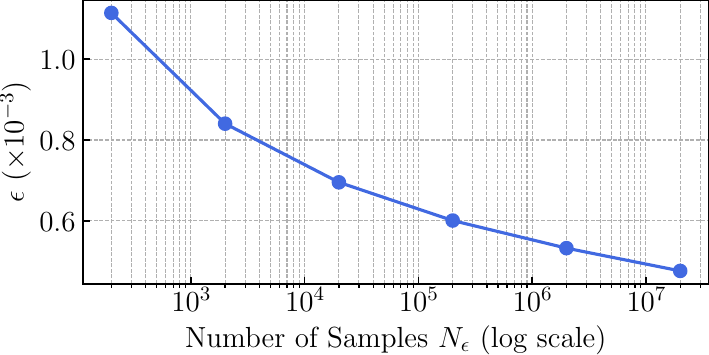}
    \caption{Illustration of sample complexity rate for the car example, showing the increase in the required number of samples as the tolerance $\epsilon$ decreases. } 
    \label{fig:epsilon_vs_Num_samples}
\end{figure}

\looseness -1 The unknown dynamics is modeled using a separate GP for each output dimension; the number of samples $\numsamples$ are computed for the vector-valued GP according to \cref{rem:vector_valued}.
Given the number of samples $N$ and the desired confidence level of $\delta = 0.01$ (corresponding to safety for all times with at least $99\%$), the sample complexity bound \eqref{eqn:sample_complexity} allows us to compute the corresponding $\epsilon$, which is the quantile of small ball probability. \cref{fig:epsilon_vs_Num_samples} shows the decay of $\epsilon$ with increasing $N$. As expected, $\epsilon$ decreases with larger $N$, however, beyond a certain $N$, the change in $\epsilon$ is small. We choose $N \in \{2 \times 10^2, 2 \times 10^4, 2 \times 10^7 \}$ and, with corresponding constants $\epsilon$, compute tightenings using \eqref{eq:ci_tightenings},
where $L_P=1.0016$ is the Lipschitz constant in the $P$-weighted norm~\eqref{eq:lip_norm} optimized as described in~\cref{sec:simulation_implementaion}. 
\cref{fig:samplingMPC_car} shows the resulting reachable set, obtained by forward sampling from the GP posterior around the given control sequence, and over-approximating it with $\epsdyn$-balls. We observe that as the number of samples increases, $\epsilon$ decreases, and we obtain a tighter over-approximation of the true reachable set.

In comparison, \cref{fig:RobustMPC_car} illustrates a reachable set obtained by the robust tube-based GP-MPC from \cite{koller_learning-based_2018}, which 
neglects dependence of $\dynGP(x_k)$ and $\dynGP(x_{k+1})$, and makes sequential over-approximations using ellipsoids. This leads to an exponential increase in the reachable set,
resulting in extremely conservative behavior already within a small horizon of $H=14$;
for larger horizon lengths, the size of the ellipsoids is too large for numerical computation. 

Note that the sampling-based reachable set is still an over-approximation of the true reachable set, which may be conservative due to the robust propagation of noise and remaining $\epsilon$ epistemic uncertainty using the Lipschitz constant. 
Theoretically, increasing $N$ can render $\epsilon$ arbitrarily small, resulting in a tighter over-approximation of the reachable set. Efficiently propagating the irreducible noise is a promising direction for future work to further reduce conservatism.

\subsection{Safe closed-loop control with pendulum dynamics}

\begin{figure}
    \centering
    \includegraphics[width=0.95\linewidth]{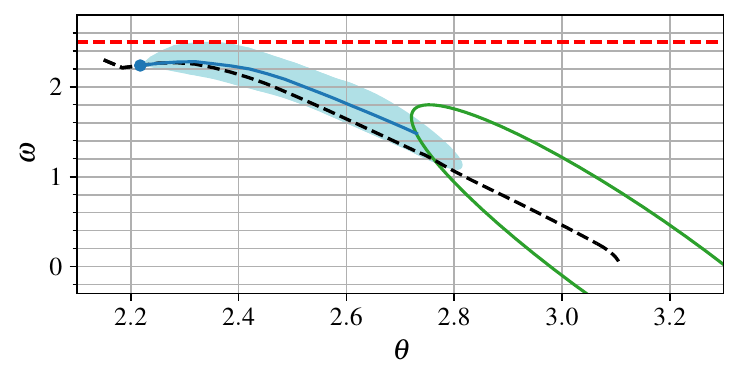}
    \caption{Demonstration of safe closed-loop control using the proposed sampling-based GP-MPC with pendulum dynamics.  
The blue dot indicates the current state of the pendulum, the cyan region shows the computed reachable set, and the blue line depicts the predicted trajectory under the mean dynamics, all at time step $\ki=3$.
The sampling-based GPMPC ensures that the reachable set satisfies the angular velocity constraint (red dashed line) and enters the terminal set (green ellipse). 
The dashed black line represents the resulting closed-loop trajectory until stabilization at the upright position.}
    \label{fig:pendulum}
\end{figure}

In this section, we demonstrate the effectiveness of the proposed sampling-based GP-MPC, using a receding-horizon implementation (\cref{alg:rhc_update_Set}), showcasing its closed-loop constraint satisfaction properties. We consider a pendulum whose nonlinear dynamics are given by:
\begin{align}
\!   \! \begin{bmatrix}\!
        \theta(\ki\!+\!1) \!\!\\
        \omega(\ki\!+\!1)\!\!
    \end{bmatrix}  \!\!=\!\!  \underbrace{\begin{bmatrix}
        \!\theta(\ki) \!+\! \omega(\ki) \Delta\!\!\\
        \omega(\ki)
    \end{bmatrix}}_{\eqqcolon \dynNom(\state,\coninput)} \!\!+ \!\! \begin{bmatrix}
        0\\
        1
    \end{bmatrix} \!
    \underbrace{\!\begin{bmatrix}
        \!\dfrac{-g_a \!\sin(\theta(\ki)) \Delta}{l} \!+\! \alpha(\ki) \Delta \!\!
    \end{bmatrix}}_{\eqqcolon \dynGP(\state, \coninput)} \label{eqn:dyn_pendulum} 
\end{align}
The pendulum state is $\state = [\theta, \omega]^\top$, where $\theta[\si{rad}]$ denotes the angular position, $\omega[\si{rad/s}]$, the angular velocity, and the control input is the angular acceleration $\alpha[\si{rad/s^2}]$. 
The constant parameter $l = 10[\si{m}]$ denotes the length of the pendulum, $\Delta=0.015 [\si{s}]$, the discretization time and $g_a=9.81 [\si{m/s^2}]$, the acceleration due to gravity.
The unknown part, \mbox{$\dynGP: \R^2 \to \R$}, \mbox{$\dynGP(\state, \coninput) = \dynGP(\theta, \alpha)$}, is a nonlinear function of angular position and acceleration.
The GP is trained using $D = 36$ prior data points on an equally-spaced
$4 \times 9$ mesh grid in the constraint set \mbox{$(\theta, \alpha) = \{ [2.1,3.6] \times[-5,5] \}$}. Starting from $(\theta, \omega) = (2.15, 2.3)$, the task is to stabilize in upright position: $(\theta, \omega) = (\pi, 0)$.

To design a terminal in gradients (\cref{assump:terminal_cost_lyap}), we find a common quadratic Lyapunov function with a linear terminal controller that stabilizes all linearized dynamics in the terminal set, which is computed by sampling 100 dynamics from the GP posterior around the upright position. 
We use the same controller in prediction, and compute the Lipschitz constant $L_P=0.96$ using a $P$-weighted norm induced by the Lyapunov function and the terminal controller as described in \eqref{eq:lip_norm}.

We fix tolerance $\epsilon=2\times 10^{-3}$ and approximate the corresponding small ball probability ($\phi(\epsilon)$). Setting $\delta=10^{-3}$ to ensure safety with at least 99.9\% and using bounded noise $\noisebound = 10^{-4}$, we compute the required number of samples \eqref{eqn:sample_complexity_bounded_noise} and find that $N=70$ samples are sufficient.


\cref{fig:pendulum} shows the closed-loop trajectory of the pendulum along with the reachable set at $k=3$. The proposed sampling-based GP-MPC can stabilize the pendulum at the upright position while satisfying all state and input constraints despite the uncertain dynamics. The MPC is solved with real-time iteration, which runs a single SQP iteration per time step. On an AMD EPYC 7543P Processor (2.80GHz), with GP sampling parallelized on an RTX 4090 GPU, a single SQP iteration takes 104.6 $\pm$ 11.9 ms. 

\section{Conclusion}
In this work, we proposed a framework for efficiently propagating epistemic uncertainty using a finite number of dynamics samples drawn from a GP posterior. 
We introduced a novel sample complexity result that bounds the required number of sampled functions and enables the construction of a reachable set that contains the true trajectory with high probability, by simulating only a finite number of samples. 
Building on this, we proposed a sampling-based GP-MPC scheme that is recursively feasible by removing falsified samples during closed-loop operation and provides theoretical guarantees on closed-loop safety and stability.
By establishing rigorous theoretical guarantees, this work opens up many possibilities for the safe and reliable deployment of sampling-based methods in real-world applications.
For future work, it would be interesting to further reduce conservatism by efficient propagation of residual $\epsilon$-epistemic uncertainty and aleatoric noise.


\bibliographystyle{unsrt}
\bibliography{ref}           

\appendix
\section{Sample complexity proofs}
\label{apx:aux_lemma}

The following lemma provides high probability error bounds used prove an upper bound on $\RKHSdatanorm[\dynGPtrue -\mu]^2/2$ in \cref{lem:C_D_norm_bound}.
\begin{lemma}[\!\!{\cite[\!Theorem 3.11]{abbasi2013online}}]
    \label{thm: beta}  \looseness -1 
    \!\!Let \cref{assump:q_RKHS} hold and  $\sqrt{\betadata[D]} \!=\! \Bg \!+\! \sqrt{ \ln(\det(\identity \!+\! \lambda^{-2} K_D) + 2\ln(2/\delta)}$.
    Then, for all \mbox{$\gpinp \in \mathcal{Z}$}, with probability at least $1-\delta/2$,
    it holds that  
    \begin{align*}
        |\dynGPtrue(\gpinp) - \muconst(\gpinp)| \leq \sqrt{\betadata[D]} \gppostvar (\gpinp).
    \end{align*}
\end{lemma}

Also, \cref{lem:C_D_norm_bound} relates the prior RKHS norm with the posterior norm for any function $\dynGP \in \RKHS$ using \cite[Appendix B]{beta_srinivas}: 
\begin{align} \label{eq:rkhsnorm_equivalence}
    \RKHSdatanorm[\dynGP]^2 = \RKHSnorm[\dynGP]^2 + \lambda^{-2} \sum\nolimits_{i=1}^{D} \dynGP(z_i)^2. 
\end{align}

\begin{proof}[Proof of \cref{lem:C_D_norm_bound}]
    \begin{align*}
        &\RKHSdatanorm[\dynGPtrue -\mu]^2 \\ 
&\stackrel{\eqref{eq:rkhsnorm_equivalence}}{=}\RKHSnorm[\dynGPtrue -\mu]^2 + \lambda^{-2} \sum\limits_{i=1}^{D} (\dynGPtrue(z_i) - \mu(z_i))^2  \\
        &\leq \langle \dynGPtrue - \mu, \dynGPtrue - \mu\rangle_k + \lambda^{-2}\betadata[D] \sum\limits_{i=1}^{D} \sigma(z_i)^2 \\
        &\labelrel={step:kernel_coeff} \RKHSnorm[\dynGPtrue]^2 \!+\! \RKHSnorm[\mu]^2 - 2\langle \dynGPtrue,\! \sum\limits_{i=1}^{D} \alpha_i k(\cdot, z_i)\rangle_k \!+\! \lambda^{-2}\betadata[D]\!\! \sum\limits_{i=1}^{D} \!\sigma(z_i)^2\\
        &\labelrel={step:linear_inner_prod} \RKHSnorm[\dynGPtrue]^2 \!+\! \RKHSnorm[\mu]^2 \!- 2\!\sum\limits_{i=1}^{D} \!\alpha_i \langle \dynGPtrue,  k(\cdot, z_i)\rangle_k \!+\! \lambda^{-2}\betadata[D] \!\! \sum\limits_{i=1}^{D} \! \sigma(z_i)^2 \\
        &\labelrel={step:unknown_rkhs} \RKHSnorm[\dynGPtrue]^2 + \RKHSnorm[\mu]^2 \!- 2 \sum\limits_{i=1}^{D} \alpha_i \dynGPtrue(z_i) + \lambda^{-2}\betadata[D] \! \sum\limits_{i=1}^{D} \! \sigma(z_i)^2 \\
        &\labelrel\leq{step:norm_bound} \Bg^2 + \RKHSnorm[\mu]^2 - 2\sum\limits_{i=1}^{D} \left( \alpha_i \mu(z_i) - |\alpha_i| \sqrt{\betadata[D]}\sigma(z_i)\right) \\
        & \qquad\qquad \quad \qquad \qquad \qquad \qquad \qquad  + \lambda^{-2}\betadata[D] \sum\limits_{i=1}^{D} \sigma(z_i)^2 \\
        &\labelrel={step:mu_struct} \!\!\Bg^2 + \RKHSnorm[\mu]^2 - 2\sum\limits_{i=1}^{D} \!\left( \!\alpha_i\! \sum\limits_{j=1}^{D} \!\alpha_j k(z_i,z_j) \!-\! |\alpha_i| \sqrt{\betadata[D]}\sigma(z_i)\!\right) \\
        & \qquad\qquad \quad \qquad \qquad \qquad \qquad \qquad  + \lambda^{-2}\betadata[D] \sum\limits_{i=1}^{D} \sigma(z_i)^2 \\
        &\labelrel={step:mu_struct_reproducing} \Bg^2 - \RKHSnorm[\mu]^2 + 2 \sum\limits_{i=1}^{D}|\alpha_i| \sqrt{\betadata[D]}\sigma(z_i) +  \lambda^{-2}\betadata[D] \sum\limits_{i=1}^{D} \sigma(z_i)^2 \\
        &\eqqcolon
        2 C_D .
    \end{align*}
The first equality follows from \eqref{eq:rkhsnorm_equivalence}. Both inequalities follow using \cref{thm: beta}. Step \eqref{step:kernel_coeff} 
follows by the definition of the GP posterior mean \eqref{eq:mean_update}. Step \eqref{step:linear_inner_prod} follows from the linearity of the inner product. 
Step \eqref{step:unknown_rkhs} and \eqref{step:norm_bound} follow respectively by reproducing property since $\dynGPtrue \in \RKHS$ and bounded norm $\RKHSnorm[\dynGPtrue]\leq \Bg$ (\cref{assump:q_RKHS}).  
Finally, steps \eqref{step:mu_struct} and \eqref{step:mu_struct_reproducing} follow respectively from the definition of mean and the reproducing property since $\mu \in \RKHS$, particularly, 
\begin{align*}
    \RKHSnorm[\mu]^2 = \langle \mu, \mu \rangle_k &= \Big\langle \sum\limits_{i=1}^{D} \!\alpha_i k(z_i,\cdot), \sum\limits_{j=1}^{D} \!\alpha_j k(\cdot,z_j) \Big\rangle_k \\
    &= \sum\limits_{i=1}^{D} \alpha_i\! \sum\limits_{j=1}^{D} \!\alpha_j k(z_i,z_j). \quad\quad\quad\qedhere
\end{align*}
\end{proof}

The following lemma relates the small ball probability for prior and posterior kernels.
\begin{lemma}[\!{\cite[Corollary 4]{anderson1955integral}}]\label{lem:small_ball_comparison_prior_posterior_variance}
    Let $\dynGP_k \sim\mathcal{GP}(0, k)$ and $\dynGP_{k_D} \sim\mathcal{GP}(\mu_D, k_D)$. Then for all $\epsilon>0$,
    \begin{align*}
        \probability{\inftynorm[\dynGP_{k_D} - \mu_D] \leq \epsilon} \geq \probability{\inftynorm[\dynGP_k] \leq \epsilon}. 
    \end{align*}
\end{lemma}
The above lemma shows that given more data, the samples are more concentrated around the mean.

\subsection{Sample complexity under bounded noise} 
\label{sec:sample_complexity_bounded_noise}

\begin{lemma}\label{lem:bounded_noise_Cd}
Let \cref{assump:q_RKHS} hold and suppose that the measurement noise generating the data set $\mathcal{D}$ is bounded by $\noisebound$, i.e., $|w_i|\leq \noisebound$, $i\in\Intrange{1}{D}$.
Then, there exists a known constant 
\begin{multline*}
    C_D \coloneqq \big( \Bg^2 + \RKHSnorm[\mu]^2 - 2\sum\nolimits_{i=1}^{D} \big( \alpha_i y_i - |\alpha_i|  \noisebound\big)  \\ 
    + \lambda^{-2} \sum\nolimits_{i=1}^{D} (|y_i-\mu_i| + \noisebound)^2 \big)/ 2 \geq 0
\end{multline*}
such that $\RKHSdatanorm[\dynGPtrue -\mu]^2/2 \leq C_D$. 
\end{lemma}
\begin{proof}
    Analogous until step \eqref{step:unknown_rkhs} of \cref{lem:C_D_norm_bound}'s proof:
    \begin{align*}
        &\RKHSdatanorm[\dynGPtrue -\mu]^2 \\ 
        &\leq \RKHSnorm[\dynGPtrue]^2 + \RKHSnorm[\mu]^2 - 2\sum\nolimits_{i=1}^{D} \alpha_i \dynGPtrue(z_i) \\ 
    &\qquad + \lambda^{-2} \sum\nolimits_{i=1}^{D} (\dynGPtrue(z_i) - \mu(z_i))^2 \numberthis\\
    &\leq \Bg^2 + \RKHSnorm[\mu]^2 - 2\sum\nolimits_{i=1}^{D} (\alpha_i y_i - |\alpha_i|\noisebound) \\ 
    & \qquad + \lambda^{-2} \sum\nolimits_{i=1}^{D} (|y_i-\mu_i| + \noisebound)^2 = 2 C_D.
\end{align*}
In the proof, the last inequality follows from $\RKHSnorm[\dynGPtrue]^2\leq \Bg$~(\cref{assump:q_RKHS}), $\dynGPtrue(z_i)=y_i-w_i$ and $|w_i|\leq \noisebound$.
\end{proof}

The following corollary presents an alternative sample complexity result under a special case of bounded measurement noise.

\begin{corollary}[Sample complexity with bounded noise] \label{coro:sample_complexity_bounded_noise}
    Let \cref{assump:q_RKHS} hold and suppose $|\noise_i| \leq \noisebound, \forall i \in \Intrange{1}{D}$. Consider any  $\epsilon>0$. Let $\numsamples \in \Nat$ be the total number of GP samples, $\dynGP^\n \sim GP(\mu, k_D)$ with 
\begin{align}
    \numsamples \geq N_\epsilon:=\frac{\log\left(\delta\right)}{\log \left(1 - e^{- \left(C_D +\phi(\epsilon)\right)} \right) }, \label{eqn:sample_complexity_bounded_noise}
\end{align}
where $C_D$ is a known constant defined as
\begin{multline}
    C_D= \big( \Bg^2 + \RKHSnorm[\mu]^2 - 2\sum\nolimits_{i=1}^{D} \big( \alpha_i y_i - |\alpha_i|  \noisebound\big) \\ + \lambda^{-2} \sum\nolimits_{i=1}^{D} (|y_i-\mu_i| + \noisebound)^2 \big)/ 2.
\end{multline}
Then $\exists \n \in \Intrange{1}{\numsamples}$ such that 
$ \inftynorm[\dynGP^\n - \dynGPtrue] < \epsilon$ with probability at least $1-\delta$.
\end{corollary}
\begin{proof}
Given that the noise is bounded, the upper bound on $\RKHSdatanorm[\dynGPtrue -\mu]^2/2 \leq C_D$ holds deterministically (see \cref{lem:bounded_noise_Cd}). Thus, \eqref{eq:sample_prob} also holds deterministically. Since no joint event needs to be satisfied, choosing $N$ as in \eqref{eqn:sample_complexity_bounded_noise} (see $\log \delta$ term) ensures that there exists at least one $\n \in \Intrange{1}{\numsamples}$ such that $\probability{ \|\dynGP^\n - \dynGPtrue\|_{\infty} < \epsilon} \geq 1 - \delta$.
\end{proof}
\cref{coro:sample_complexity_bounded_noise} differs from \cref{thm:sample_complexity} in its assumption on measurement noise. \cref{thm:sample_complexity} applies to more general, conditionally $\lambda$-sub-Gaussian noise, which requires identifying a suitable variance proxy $\lambda$ in practice. In contrast, \cref{coro:sample_complexity_bounded_noise} requires a known upper bound on the noise measurement.
In this case, the bound $C_D$ holds deterministically using the assumed bound on the noise, resulting in sample complexity that scales with $\log (1/\delta)$ instead of $\log (2/\delta)$ in \cref{thm:sample_complexity}.

\section{Stability proofs}
\label{apx:stability}
The following lemma provides a bound on the one-step decrease of the optimal cost, which is leveraged in the proof of Theorem~\ref{thm:stability}.
\begin{lemma} \label{lem:cost_decrease} Let \cref{assump:q_RKHS,assump:lipschitz_dyn,assump:bounded_noise,assum:safeset,assump:lipschitz_cost,assump:terminal_cost_lyap} hold and suppose that Problem~\eqref{eq:SafeGPMPCinf} is feasible at time $\ki\in\mathbb{N}$. 
    The optimal cost $J^\star(\cdot, \cdot)$  satisfies 
\begin{multline*}
    J^\star(\state(\ki+1), \N_{\ki+1}) - J^\star(\state(\ki), \N_{\ki}) \\ 
    \leq |\N_{\ki+1}| \left(\Ljc + \stagecostNA_s  - \stagecost{\state(\ki)}{\coninput(\ki)}\right).
\end{multline*}
where $\Ljc = K_1 \noisebound 
+ K_2 \epsilon$ with constants $K_1, K_2>0$.
\end{lemma}
\begin{proof}
    Using the feasible candidate solution from the proof of \cref{thm:safety_recursive_feasibility} with the terminal input $\coninput_{\horizon-1|\ki+1} \coloneqq \coninput^\star_{\horizon|\ki} = u_{\mathrm{f}}$, we obtain
\begin{align*}
& J^\star(\state(\ki+1), \N_{\ki+1}) - J^\star(\state(\ki), \N_{\ki})  \\
&\leq \sum_{\n \in \N_{\ki+1}} \left(\sum_{i=0}^{\horizon - 1} \stagecost{\state^\n_{i|\ki+1}}{\coninput^\star_{i+1|\ki}} + \terminalcost{\state^\n_{\horizon|\ki+1}} \right) \\
& \qquad - \sum_{\n \in \N_{\ki}} \!\!\!\left(\sum_{i=1}^{\horizon - 1} \!\stagecost{\state^\n_{i|\ki}}{\coninput^\star_{i|\ki}} +\stagecost{\state(\ki)}{\coninput(\ki)} + \terminalcost{\state^\n_{\horizon|\ki}} \!\right)\\
&\labelrel\leq{step:filter_dyn} \sum_{\n \in \N_{\ki+1}} \!\!\left(\sum_{i=0}^{\horizon - 1} \stagecost{\state^\n_{i|\ki+1}}{\coninput^\star_{i+1|\ki}} + \terminalcost{\state^\n_{\horizon|\ki+1}} \right) \\
& \qquad - \sum_{\n \in \N_{\ki+1}} \!\!\!\!\left(\sum_{i=1}^{\horizon - 1} \!\stagecost{\state^\n_{i|\ki}}{\coninput^\star_{i|\ki}} +\stagecost{\state(\ki)}{\coninput(\ki)} + \terminalcost{\state^\n_{\horizon|\ki}} \!\!\right)\\
&= \!\!\!\!\sum_{\n \in \N_{\ki+1}} \!\!\!\!\Bigg(\sum_{i=0}^{\horizon - 2} \!\!\stagecost{\state^\n_{i|\ki+1}\!}{\!\coninput^\star_{i+1|\ki}\!} \!+\! \stagecost{\state^\n_{\horizon-1|\ki+1}\!}{\!\terminalinput\!} \!+\! \terminalcost{\state^\n_{\horizon|\ki+1}} \\
& \; \; -\!\! \sum_{i=0}^{\horizon - 2} \!\stagecost{\state^\n_{i+1|\ki}}{\!\coninput^\star_{i+1|\ki}} \!+\!\stagecost{\state(\ki)}{\!\coninput(\ki)} \!+\! \terminalcost{\state^\n_{\horizon|\ki}} \!\!\Bigg)\!.
\!\!\! \label{eqn:cost_diff} \numberthis
\end{align*}
Step \eqref{step:filter_dyn} follows since  $\stagecost{x}{u} \geq 0, \terminalcost{x} \geq 0, \forall(x,u)\in \mathcal{Z}$ and the set $\N_k$ is non-increasing by construction~\eqref{eqn:adaptive_dyn_set}.
Using \cref{assump:lipschitz_cost}, for $i\in\Intrange{0}{\horizon-2}$ it holds that
\begin{align*}
    \stagecost{\state^\n_{i|\ki+1}}{\coninput^\star_{i+1|\ki}} - \stagecost{\state^\n_{i+1|\ki}}{\coninput^\star_{i+1|\ki}} &\leq \Lipstagecost \euclidnorm[\state^\n_{i|\ki+1} - \state^\n_{i+1|\ki}]\\
    &\stackrel{\ref{eqn:adaptive_dyn_set}}{\leq} \Lipstagecost c_i,
\end{align*}
and on substituting it in \cref{eqn:cost_diff}, we get,
\begin{align*}
&\leq \sum_{\n \in \N_{\ki+1}} \Bigg( \sum_{i=0}^{\horizon - 2}  \Lipstagecost c_i +\! \stagecost{\state^\n_{\horizon-1|\ki+1}}{\terminalinput}  \\
& \quad + \terminalcost{\state^\n_{H|\ki+1}} - \terminalcost{\state^\n_{H|\ki}}\! - \stagecost{\state(\ki)}{\coninput(\ki)} \Bigg)  \numberthis \label{eq:substitute_c_i}
\end{align*}
and similarly using \cref{assump:lipschitz_cost} for terminal cost,
\begin{align*}
    &\terminalcost{\state^\n_{H|\ki+1}} - \terminalcost{\state^\n_{H|\ki}} \\
    &\leq \terminalcost{\state^\n_{H|\ki+1}} - \terminalcost{\state^\n_{H-1|\ki+1}} + \Liptermcost \euclidnorm[\state^\n_{H-1|\ki+1} - \state^\n_{H|\ki}]\\
    &\stackrel{\ref{eqn:adaptive_dyn_set}}{\leq} \terminalcost{\state^\n_{H|\ki+1}} - \terminalcost{\state^\n_{H-1|\ki+1}} + \Liptermcost c_{\horizon-1},
\end{align*}
we get the following from \cref{eq:substitute_c_i},
\begin{align*}
&\leq \sum_{\n \in \N_{\ki+1}} \Bigg( \sum_{i=0}^{\horizon - 2}  \Lipstagecost c_i + \Liptermcost c_{\horizon-1} - \stagecost{\state(\ki)}{\coninput(\ki)} \\
& \quad +  \terminalcost{\state^\n_{H|\ki+1}} \!-\! \terminalcost{\state^\n_{H-1|\ki+1}} \!+\! \stagecost{\state^\n_{\horizon-1|\ki+1}}{\terminalinput} 
\Bigg)\\
&\leq  |\N_{\ki+1}| \left( K_1 \bar{\noise} + K_2 \epsilon + \stagecostNA_s  - \stagecost{\state(\ki)}{\coninput(\ki)}\right),
\end{align*}
where $K_1 = \|\gpsubspace\|( \sum_{i=0}^{\horizon-2} \Lipdyn^i \Lipstagecost  + \Lipdyn^{\horizon-1}  \Liptermcost)$ and $K_2\! =\! \|\gpsubspace\|( \Lipstagecost (\sum_{i=0}^{\horizon-2} \!\Lipdyn^i  +  2 \! \sum_{j=0}^{i-1} \!\Lipdyn^j)  + \Liptermcost(\Lipdyn^{\horizon-1} \! +  2\!\sum_{j=0}^{\horizon-2} \!\Lipdyn^j ))$ are constants. The last inequality follows using \cref{assump:terminal_cost_lyap} and the definition of $c_i$ (c.f. \cref{lem:tightening}). Using $\Ljc = K_1 \noisebound + K_2 \epsilon$ concludes the proof.
\end{proof}
\begin{proof}[Proof of \cref{thm:stability}]
Using \cref{lem:cost_decrease} and 
using telescopic sum until some finite time T, we obtain
\begin{align*}
J^\star(\state(T),& \ \N_T) -             J^\star(\state(0), \N_0) \\ 
&\leq \sum_{k=0}^{T-1} |\N_{\ki+1}| \left(\Ljc + \stagecostNA_s  - \stagecost{\state(\ki)}{\coninput(\ki)}\right) \\
&\leq |\N_{1}|  \left( T \Ljc + T \stagecostNA_s  - \sum_{k=0}^{T-1} \stagecost{\state(\ki)}{\coninput(\ki)}\right). 
\end{align*}
This implies that
\begin{align*}
\frac{1}{T} \sum_{\ki=0}^{T-1} & \stagecost{\state(\ki)}{\coninput(\ki)} \\ 
& \leq \Ljc + \stagecostNA_s  - \frac{J^\star(\state(T), \N_T) -   J^\star(\state(0), \N_0)}{T |\N_1|}.  
\end{align*}
Since the optimal cost $J^\star$ is bounded, by applying the asymptotic limit on the average cost, we obtain \cref{eqn:cost_convergence}.

To establish practical asymptotic stability (cf.~\eqref{eq:practical_stability}), we show that the optimal cost $J^\star$ is a practical Lyapunov function according to~\cite[Prop.~4.3]{faulwasser2018economic}. 
First note that the positive definite cost and the terminal cost decrease (cf. \cref{assump:terminal_cost_lyap}) ensure that $J^\star(x(k),\mathcal{N}_k)\leq |\mathcal{N}_0|V_f(x(k))$ for all $x(k)\in\mathcal{X}_f$. 
Furthermore, given that the origin is in the interior of $\mathcal{X}_f$, $V_f$ is locally upper bounded, and $\mathcal{X}$ is compact, there exists $\alpha_3\in\mathcal{K}_\infty$ satisfying   $ J^\star(\|x\|, \N) \leq \alpha_3(\|x\|)$~\cite[Sec 2.4.2]{rawlingsModelPredictiveControl2020}. 
This upper bound in combination with the decrease condition from \cref{lem:cost_decrease} ensures that $J^\star$ satisfies the conditions for a practical Lyapunov function in~\cite[Prop.~4.3]{faulwasser2018economic}.
\end{proof}

\end{document}